\documentclass[12pt]{article}
\usepackage{amsmath,amssymb}
\usepackage{bm}
\usepackage{color}
\usepackage{cite}
\usepackage{mathtools}
\usepackage{here}

\usepackage{multirow}

\begin{document}

\title{
\begin{flushright}
\ \\*[-80pt]
\begin{minipage}{0.2\linewidth}
\normalsize
EPHOU-23-001\\*[50pt]
\end{minipage}
\end{flushright}
{\Large \bf
Quark hierarchical structures in modular symmetric flavor models at level 6
\\*[20pt]}}

\author{
Shota Kikuchi,
~Tatsuo Kobayashi,
~Kaito Nasu,
\\~Shohei Takada, and
~Hikaru Uchida
\\*[20pt]
\centerline{
\begin{minipage}{\linewidth}
\begin{center}
{\it \normalsize
Department of Physics, Hokkaido University, Sapporo 060-0810, Japan} \\*[5pt]
\end{center}
\end{minipage}}
\\*[50pt]}

\date{
\centerline{\small \bf Abstract}
\begin{minipage}{0.9\linewidth}
\medskip
\medskip
\small
We study modular symmetric quark flavor models 
without fine-tuning.
Mass matrices are written in terms of modular forms, and 
modular forms in the vicinity of the modular fixed points become hierarchical depending on their residual charges.
Thus modular symmetric flavor models 
 in the vicinity of the modular fixed points have a possibility to describe mass hierarchies without fine-tuning.
Since describing quark hierarchies without fine-tuning requires $Z_n$ residual symmetry with $n\geq 6$, we focus on $\Gamma_6$ modular symmetry in the vicinity of the cusp $\tau=i\infty$ where $Z_6$ residual symmetry remains.
We use only modular forms belonging to singlet representations of $\Gamma_6$ to make our analysis simple.
Consequently, viable quark flavor models are obtained without fine-tuning.
\end{minipage}
}

\begin{titlepage}
\maketitle
\thispagestyle{empty}
\end{titlepage}

\newpage


\section{Introduction}
\label{Intro}

The origin of quark and lepton flavor structures such as hierarchical masses and mixing angles is one of challenging issues in particle physics.
Indeed, many works were done in order to solve the problem.
Among such works, modular symmetric flavor models are interesting.
In these flavor models, the quark and lepton mass matrices are written in terms of modular forms, which are holomorphic functions of the modulus $\tau$ \cite{Feruglio:2017spp} \footnote{The modular flavor symmetry was also studied from the top-down approach such as string theory 
\cite{Ferrara:1989bc,Ferrara:1989qb,Lerche:1989cs,Lauer:1990tm,Kobayashi:2018rad,Kobayashi:2018bff,Ohki:2020bpo,Kikuchi:2020frp,Kikuchi:2020nxn,
Kikuchi:2021ogn,Almumin:2021fbk,Baur:2019iai,Nilles:2020kgo,Baur:2020jwc,Nilles:2020gvu}.}.
It is well known that the finite modular groups $\Gamma_N$ for $N=2,~3,~4,~5$ are isomorphic to the non-Abelian finite groups $S_3$, $A_4$, $S_4$ and $A_5$, respectively \cite{deAdelhartToorop:2011re}.
This is interesting since the non-Abelian finite groups are long familiar in flavor models for quarks and leptons \cite{Altarelli:2010gt,Ishimori:2010au,Ishimori:2012zz,Kobayashi:2022moq,Hernandez:2012ra,King:2013eh,King:2014nza,Tanimoto:2015nfa,King:2017guk,Petcov:2017ggy,Feruglio:2019ktm}.
Inspired by this point, the modular symmetric lepton flavor models have been proposed in $\Gamma_2\simeq S_3$ \cite{Kobayashi:2018vbk}, $\Gamma_3\simeq A_4$ \cite{Feruglio:2017spp}, $\Gamma_4\simeq S_4$ \cite{Penedo:2018nmg} and $\Gamma_5\simeq A_5$ \cite{Novichkov:2018nkm,Ding:2019xna}.
In addition, modular symmetries at levels $6$ \cite{Li:2021buv} and $7$ \cite{Ding:2020msi} were studied.
Furthermore, modular forms of other groups were also studied \cite{Kobayashi:2018bff,Liu:2019khw,Novichkov:2020eep,Liu:2020akv,Liu:2020msy}.

Using these various modular forms, the mass ratios and flavor mixing of quarks and leptons have been discussed successfully in these years.
Phenomenological studies have been developed in many works and interesting results have been obtained \cite{
Criado:2018thu,
Kobayashi:2018scp,
Ding:2019zxk,
Novichkov:2018ovf,
Kobayashi:2019mna,Wang:2019ovr,
Chen:2020udk,
deMedeirosVarzielas:2019cyj,
  	Asaka:2019vev,
Asaka:2020tmo,deAnda:2018ecu,Kobayashi:2019rzp,Novichkov:2018yse,Kobayashi:2018wkl,Okada:2018yrn,Okada:2019uoy,Nomura:2019jxj, Okada:2019xqk,
  	Nomura:2019yft,Nomura:2019lnr,Criado:2019tzk,
  	King:2019vhv,Gui-JunDing:2019wap,deMedeirosVarzielas:2020kji,Zhang:2019ngf,Nomura:2019xsb,Kobayashi:2019gtp,Lu:2019vgm,Wang:2019xbo,King:2020qaj,Abbas:2020qzc,Okada:2020oxh,Okada:2020dmb,Ding:2020yen,Okada:2020rjb,Okada:2020ukr,Nagao:2020azf,Wang:2020lxk,
  	Okada:2020brs,Yao:2020qyy}.
However, one needs to  fine-tune coefficients of modular forms in Yukawa couplings in order 
to describe the hierarchical structure of fermion masses, 
in particular quark mass hierarchies.


Describing the lepton flavors without fine-tuning on  modular invariant models was proposed in Ref.~\cite{Novichkov:2021evw}.
Authors focused on the vicinity of three modular fixed points, $\tau=i$, $\omega$ $(=e^{2\pi i/3})$ and $i\infty$ where  residual symmetries remain \cite{Novichkov:2018ovf}.
The values of  modular forms become hierarchical as close to these modular fixed points due to approximate residual symmetries.
Then, the hierarchy among  values of the modular forms  is determined by charges of  residual symmetries at the modular fixed points.
For instance the modular forms of $\Gamma_4\simeq S_4$ with $Z_4$ ($T$-transformation) charges 0, 1, 2 and 3 can take the sizes 1, $\varepsilon$, $\varepsilon^2$ and $\varepsilon^3$, respectively in the vicinity of $\tau=i\infty$, where  $\varepsilon$ expresses the deviation from the modular fixed points.
Indeed viable lepton models on the double covering groups of $\Gamma_N$, $\Gamma_3'\simeq A_4'$, $\Gamma_4'\simeq S_4'$ and $\Gamma_5\simeq A_5'$, were studied in Ref. \cite{Novichkov:2021evw}.
This is one successful way to generate hierarchical structures without fine-tuning.
Nevertheless the realization of the quark flavor structure is not straightforward.
Experiments show  mass hierarchies of up sector quarks as $m_u/m_t\sim10^{-5}$ and $m_c/m_t\sim10^{-2}\textrm{-}10^{-3}$ and ones of down sector quarks as $m_d/m_b\sim 10^{-3}$ and $m_s/m_b\sim 10^{-2}$ \cite{Zyla:2020zbs}.
Suppose that $\varepsilon ={\cal O}(0.1)$.
Then, we could explain these mass ratios except $m_u/m_t\sim10^{-5}$.
However,  $\varepsilon^5$ does not appear in the framework of the finite modular group of level $N$ less than 6 since the present residual symmetries $Z_2$, $Z_3$ and $Z_N$ at $\tau=i$, $\omega$ and $i\infty$ do not possess the charge larger than 4.
Thus describing the quark flavor structure without fine-tuning requires the way generating hierarchical mass ratios including $\varepsilon^5={\cal O}(10^{-5})$.

One way is to relax the quark mass eigenvalues by tuning the values of coupling constants in Yukawa couplings.
In Ref.~\cite{Petcov:2022fjf}, the quark flavor model with $A_4$ modular symmetry was studied and succeeded to generate both up and down sector quark mass hierarchies by adjusting one coupling constant ratio denoted by $g_u/g_d$ to ${\cal O}(10)$.
As a result, quark mass hierarchies originate from following two origins,
\begin{enumerate}
  \renewcommand{\labelenumi}{(\roman{enumi})}
  \item{The vacuum expectation value (VEV) of the modulus $\tau$ (the deviation from the modular fixed points),}
  \item{The coupling constants in Yukawa couplings.}
\end{enumerate}

Another way is to introduce the finite modular symmetry including $Z_n$ residual symmetry with $n\geq 6$.
In such models, the modular forms in the vicinity of the symmetric points can take the sizes 1, $\varepsilon$, $\cdots$, $\varepsilon^{n-1}$ depending on their residual charges.
Hence, it may be possible to generate mass hierarchies in both up and down sector quarks  without fine-tuning using the hierarchical values of the modular forms up to $\varepsilon^5$.
Note that in this way quark mass hierarchies simply originate from (i) above.
In this paper, we study the modular symmetric quark flavor model with the finite modular group of level 6, $\Gamma_6\simeq S_3\times A_4$.
The finite modular symmetry $\Gamma_6$ breaks into $Z_6$ ($T$-transformation) residual symmetry at $\tau=i\infty$ and can generate the hierarchical values of the modular forms up to $\varepsilon^5$ in the vicinity of $\tau=i\infty$.

This paper is organized as follows.
In section \ref{sec:2}, we study generic aspects for  the modular symmetric quark flavor models being able to realize both up and down sector quark mass hierarchies without fine-tuning along the lines proposed in Ref.~\cite{Novichkov:2021evw}.
In section \ref{sec:3}, we study quark flavor models with the finite modular group $\Gamma_6$.
Section \ref{sec:4} is our conclusion.
We summarize group theoretical aspects of $\Gamma_6$ in appendix \ref{appendix:A} and the modular forms of level 6 in appendix \ref{appendix:B}.


\section{Hierarchical quark mass matrices without fine-tuning}
\label{sec:2}

In this section, we present modular symmetric quark flavor models without fine-tuning.
We start from the  following assignment of modular weights to supermultiplets:
\begin{itemize}
  \item{quark doublets $Q=(Q^1, Q^2, Q^3)$ are assigned into three-dimensional (reducible or irreducible ) representation of a finite modular group with weight $-k_Q$,}
  \item{up sector quark singlets $u_R=(u_R^1, u_R^2, u_R^3)$ are assigned into three-dimensional (reducible or irreducible ) representation of a finite modular group with weight $-k_u$,}
  \item{down sector quark singlets $d_R=(d_R^1, d_R^2, d_R^3)$ are assigned into three-dimensional (reducible or irreducible ) representation of a finite modular group with weight $-k_d$,}
  \item{each of up and down sector Higgs fields $H_{u,d}$ is assigned into one-dimensional representations of a finite modular group with weight $-k_{H_{u,d}}$.}
\end{itemize}
Note that three-dimensional representations are constructed by combining singlets, doublets and triplets of any finite modular groups.
The most general form of the superpotential relevant to up sector quark masses is written as
\begin{align}
  W_u = \sum_{\bm{r}_i} \left[
  Y_{\bm{r}_i}^{(k_{Y_u})}
  \begin{pmatrix}
    Q^1 & Q^2 & Q^3 \\
  \end{pmatrix}
  \begin{pmatrix}
    \alpha^{11}_{\bm{r}_i} & \alpha^{12}_{\bm{r}_i} & \alpha^{13}_{\bm{r}_i} \\
    \alpha^{21}_{\bm{r}_i} & \alpha^{22}_{\bm{r}_i} & \alpha^{23}_{\bm{r}_i} \\
    \alpha^{31}_{\bm{r}_i} & \alpha^{32}_{\bm{r}_i} & \alpha^{33}_{\bm{r}_i} \\
  \end{pmatrix}
  \begin{pmatrix}
    u_R^1 \\
    u_R^2 \\
    u_R^3 \\
  \end{pmatrix}
  H_u
  \right]_{\bm{1}},
\end{align}
where $Y_{\bm{r}_i}^{(k_{Y_u})}$ denotes the modular forms of irreducible representation $\bm{r}_i$ for weight $k_{Y_u}=k_Q+k_u+k_{H_u}$.
Some of coupling constants $\alpha^{ij}$ may be related each other when quark doublets $Q$ and/or up sector quark singlets $u_R$ belong to multiplets.
Similarly the superpotential relevant to down sector quark masses is written as
\begin{align}
  W_d = \sum_{\bm{r}_i} \left[
  Y_{\bm{r}_i}^{(k_{Y_d})}
  \begin{pmatrix}
    Q^1 & Q^2 & Q^3 \\
  \end{pmatrix}
  \begin{pmatrix}
    \beta^{11}_{\bm{r}_i} & \beta^{12}_{\bm{r}_i} & \beta^{13}_{\bm{r}_i} \\
    \beta^{21}_{\bm{r}_i} & \beta^{22}_{\bm{r}_i} & \beta^{23}_{\bm{r}_i} \\
    \beta^{31}_{\bm{r}_i} & \beta^{32}_{\bm{r}_i} & \beta^{33}_{\bm{r}_i} \\
  \end{pmatrix}
  \begin{pmatrix}
    d_R^1 \\
    d_R^2 \\
    d_R^3 \\
  \end{pmatrix}
  H_d
  \right]_{\bm{1}},
\end{align}
with $k_{Y_d}=k_Q+k_d+k_{H_d}$.
They lead to the up and down sector quark mass matrices, $M_u$ and $M_d$,
\begin{align}
  &\begin{pmatrix}
    Q^1 & Q^2 & Q^3 \\
  \end{pmatrix}
  M_u
  \begin{pmatrix}
    u_R^1 \\
    u_R^2 \\
    u_R^3 \\
  \end{pmatrix}
  =
  \sum_{\bm{r}_i} \left[
  Y_{\bm{r}_i}^{(k_{Y_u})}
  \begin{pmatrix}
    Q^1 & Q^2 & Q^3 \\
  \end{pmatrix}
  \begin{pmatrix}
    \alpha^{11}_{\bm{r}_i} & \alpha^{12}_{\bm{r}_i} & \alpha^{13}_{\bm{r}_i} \\
    \alpha^{21}_{\bm{r}_i} & \alpha^{22}_{\bm{r}_i} & \alpha^{23}_{\bm{r}_i} \\
    \alpha^{31}_{\bm{r}_i} & \alpha^{32}_{\bm{r}_i} & \alpha^{33}_{\bm{r}_i} \\
  \end{pmatrix}
  \begin{pmatrix}
    u_R^1 \\
    u_R^2 \\
    u_R^3 \\
  \end{pmatrix}
  \langle H_u \rangle
  \right]_{\bm{1}}, \\
  &\begin{pmatrix}
    Q^1 & Q^2 & Q^3 \\
  \end{pmatrix}
  M_d
  \begin{pmatrix}
    d_R^1 \\
    d_R^2 \\
    d_R^3 \\
  \end{pmatrix}
  =
  \sum_{\bm{r}_i} \left[
  Y_{\bm{r}_i}^{(k_{Y_d})}
  \begin{pmatrix}
    Q^1 & Q^2 & Q^3 \\
  \end{pmatrix}
  \begin{pmatrix}
    \beta^{11}_{\bm{r}_i} & \beta^{12}_{\bm{r}_i} & \beta^{13}_{\bm{r}_i} \\
    \beta^{21}_{\bm{r}_i} & \beta^{22}_{\bm{r}_i} & \beta^{23}_{\bm{r}_i} \\
    \beta^{31}_{\bm{r}_i} & \beta^{32}_{\bm{r}_i} & \beta^{33}_{\bm{r}_i} \\
  \end{pmatrix}
  \begin{pmatrix}
    d_R^1 \\
    d_R^2 \\
    d_R^3 \\
  \end{pmatrix}
  \langle H_d \rangle
  \right]_{\bm{1}}.
\end{align}

We expect that the coefficients $\alpha^{ij}$ and $\beta^{ij}$ are of ${\cal O}(1)$, because 
we do not explain quark mass hierarchies by using hierarchies of these coefficients.
In particular, we restrict all coupling constants $\alpha^{ij}$ and $\beta^{ij}$ to $\pm 1$ 
and study the realization of the orders of mass ratios and the Cabibbo, Kobayashi, Maskawa (CKM) matrix elements.
Then free parameter is only the value of the modulus $\tau$ (and the choices of $+1$ or $-1$ in coupling constants $\alpha^{ij}$ and $\beta^{ij}$).

\begin{table}[H]
 \centering
  \begin{tabular}{c|c}
   \hline
   $(m_u,m_c,m_t)/m_t$ & $(1.26\times 10^{-5},7.38\times 10^{-3},1)$ \\ \hline
   $(m_d,n_s,m_b)/m_b$ & $(1.12\times 10^{-3},2.22\times 10^{-2},1)$ \\ \hline
  \end{tabular}
  \caption{Observed values of  quark masses \cite{Zyla:2020zbs}.}
  \label{tab:quark_masses}
\end{table}
In order to realize hierarchical quark masses as shown in Table \ref{tab:quark_masses} without fine-tuning, it is necessary to generate hierarchies by values of the modular forms.
Actually such hierarchical values of the modular forms can be realized in a vicinity of three modular fixed points, $\tau=i$, $\omega$ and $i\infty$.
This can be understood as follows.
As an example let us consider $Z_n$ symmetric point and quark doublets $Q$, up sector quark singlets $u_R$ and up-type Higgs field $H_u$ with the following $Z_n$ residual charges,
\begin{align}
  Q:~(1,n-1,0), \quad u_R:~(1,0,0), \quad H_u:~0.
\end{align}
Then the entities of the up sector quark mass matrix, $M_u^{ij}$, must have the following $Z_n$ residual charges to make Lagrangian modular invariant,
\begin{align}
  M_u^{ij} :~
  \begin{pmatrix}
    n-2 & n-1 & n-1 \\
    0 & 1 & 1 \\
    n-1 & 0 & 0 \\
  \end{pmatrix}.
\end{align}
In the vicinity of $Z_n$ symmetric point, the modular forms with $Z_n$ residual charge $q$, $f(\tau)$, can be expanded by the deviation from the symmetric point to the power of $q$ \cite{Novichkov:2021evw}:
\begin{enumerate}
  \item $\tau \sim i$: $f(\tau)\sim \varepsilon^q$, $\varepsilon \equiv \frac{\tau-i}{\tau+i}$,
  \item $\tau \sim \omega$: $f(\tau)\sim \varepsilon^q$, $\varepsilon \equiv \frac{\tau-\omega}{\tau-\omega^2}$,
  \item $\tau \sim i\infty$: $f(\tau)\sim \varepsilon^q$, $\varepsilon \equiv e^{-2\pi \textrm{Im}\tau/N}$ ($N$ is a level of the finite modular group).
\end{enumerate}
Thus the above up sector quark mass matrix can be evaluated as
\begin{align}
  M_u^{ij} \sim
  \begin{pmatrix}
    \varepsilon^{n-2} & \varepsilon^{n-1} & \varepsilon^{n-1} \\
    1 & \varepsilon & \varepsilon \\
    \varepsilon^{n-1} & 1 & 1 \\
  \end{pmatrix},
\end{align}
in the vicinity of $Z_n$ symmetric point.
Similarly, for the down sector quark mass matrix as well as lepton mass matrices, the modular forms take hierarchical values depending on their residual charges and lead to hierarchical mass matrices as close to the modular fixed points.
In Ref.~\cite{Novichkov:2021evw}, lepton flavor models without fine-tuning around the vicinity of the modular fixed points was studied.

On the other hand, it is difficult to realize quark mass hierarchies by the values of the modular forms in the vicinity of $\tau=i$ and $\omega$.
To realize both up and down sector quark mass hierarchies in Table \ref{tab:quark_masses} simultaneously, we may need $\varepsilon$ to the fifth power, when $\varepsilon = {\cal O}(0.1)$.
Hence we need five different residual charges.
This requirement excludes the vicinity of $\tau=i$ and $\omega$ since they correspond to $Z_2$ and $Z_3$ symmetries, respectively.
In other words, such hierarchical masses can be realized in the vicinity of the cusp $\tau=i\infty$ with $Z_N$ charge for $N\geq 6$.

Let us discuss the candidates of the modular symmetry.
As mentioned above the level of the modular symmetry must be lager than 5.
Here we focus on the levels 6 and 7, that is, $\Gamma_6\simeq S_3\times A_4$ and $\Gamma_7\simeq PSL(2,Z_7)$ as the candidates of the modular symmetry.
As the irreducible representations less than four dimension, $\Gamma_7\simeq PSL(2,Z_7)$ has only one singlet $\bm{1}$ and two triplets $\bm{3}$ and $\bm{\bar{3}}$ \cite{Ding:2020msi}; this variety of irreducible representations may not be enough 
to find the models being able to realize both up and down sector quark masses.
In contrast, $\Gamma_6\simeq S_3\times A_4$ has six singlets, $\bm{1^0_0}$, $\bm{1^0_1}$, $\bm{1^0_2}$, $\bm{1^1_0}$, $\bm{1^1_1}$ and $\bm{1^1_2}$, three doublets, $\bm{2_0}$, $\bm{2_1}$ and $\bm{2_2}$, and two triplets, $\bm{3^0}$ and $\bm{3^1}$ \cite{Li:2021buv}.
They would  be sufficient to find realistic models.
In the following, we consider the models with $\Gamma_6$ modular symmetry and realize quark flavors without fine-tuning in the vicinity of $\tau=i\infty$.


\section{The models with $\Gamma_6$ modular symmetry}
\label{sec:3}

Here we study the models with $\Gamma_6$ modular symmetry and realize the quark flavor structure without fine-tuning.
As we have mentioned in the previous section, we restrict all couplings $\alpha^{ij}$ and $\beta^{ij}$ to $\pm 1$ in quark mass matrices to avoid   fine-tuning by them.
We study the realization of the  orders of mass ratios and mixing angles.
We use only the modulus $\tau$ (and the choices of +1 or -1 in $\alpha^{ij}$ and $\beta^{ij}$) as a free parameter.

In $\Gamma_6$ modular symmetry, $\varepsilon$ to the power up to 5 can appear in mass matrices.
Indeed six $\Gamma_6$ singlets with six different $T$-charges correspond to different powers of $\varepsilon$ in the vicinity of $\tau=i\infty$ as shown in Table \ref{tab:singlets_T_ep}.
\begin{table}[H]
  \begin{center}
    \renewcommand{\arraystretch}{1.1}
    $\begin{array}{c|cccccc} \hline
      \textrm{singlet} & \bm{1^0_0} & \bm{1^1_2} & \bm{1^0_1} & \bm{1^1_0} & \bm{1^0_2} & \bm{1^1_1} \\ \hline
      T\textrm{-charge} & 0 & 1 & 2 & 3 & 4 & 5 \\ \hline
      \textrm{order} & 1 & \varepsilon & \varepsilon^2 & \varepsilon^3 & \varepsilon^4 & \varepsilon^5 \\ \hline
    \end{array}$
  \end{center}
  \caption{$T$-charges of six $\Gamma_6$ singlets and their orders in the vicinity of $\tau=i\infty$.}
  \label{tab:singlets_T_ep}
\end{table}
To realize the quark flavor structure, let us consider the following four types of mass matrices,
\begin{align}
  &\textrm{Type~I:} \quad M_u \propto
  \begin{pmatrix}
    \varepsilon^5 & \varepsilon^{3-a+b} & \varepsilon^b \\
    \pm\varepsilon^{5+a-b} & \pm\varepsilon^3 & \pm\varepsilon^a \\
    \pm\varepsilon^{5-b} & \pm\varepsilon^{3-a} & \pm1 \\
  \end{pmatrix}, \quad
  M_d \propto
  \begin{pmatrix}
    \varepsilon^3 & \varepsilon^{2-a+b} & \varepsilon^b \\
    \pm\varepsilon^{3+a-b} & \pm\varepsilon^2 & \pm\varepsilon^a \\
    \pm\varepsilon^{3-b} & \pm\varepsilon^{2-a} & \pm1 \\
  \end{pmatrix}, \label{eq:firstmassmtx} \\
  &\textrm{Type~II:} \quad M_u \propto
  \begin{pmatrix}
    \varepsilon^5 & \varepsilon^{3-a+b} & \varepsilon^b \\
    \pm\varepsilon^{5+a-b} & \pm\varepsilon^3 & \pm\varepsilon^a \\
    \pm\varepsilon^{5-b} & \pm\varepsilon^{3-a} & \pm1 \\
  \end{pmatrix}, \quad
  M_d \propto
  \begin{pmatrix}
    \varepsilon^4 & \varepsilon^{2-a+b} & \varepsilon^b \\
    \pm\varepsilon^{4+a-b} & \pm\varepsilon^2 & \pm\varepsilon^a \\
    \pm\varepsilon^{4-b} & \pm\varepsilon^{2-a} & \pm1 \\
  \end{pmatrix}, \label{eq:secondmassmtx} \\
  &\textrm{Type~III:} \quad M_u \propto
  \begin{pmatrix}
    \varepsilon^5 & \varepsilon^{2-a+b} & \varepsilon^b \\
    \pm\varepsilon^{5+a-b} & \pm\varepsilon^2 & \pm\varepsilon^a \\
    \pm\varepsilon^{5-b} & \pm\varepsilon^{2-a} & \pm1 \\
  \end{pmatrix}, \quad
  M_d \propto
  \begin{pmatrix}
    \varepsilon^3 & \varepsilon^{2-a+b} & \varepsilon^b \\
    \pm\varepsilon^{3+a-b} & \pm\varepsilon^2 & \pm\varepsilon^a \\
    \pm\varepsilon^{3-b} & \pm\varepsilon^{2-a} & \pm1 \\
  \end{pmatrix}, \label{eq:thirdmassmtx} \\
  &\textrm{Type~IV:} \quad M_u \propto
  \begin{pmatrix}
    \varepsilon^5 & \varepsilon^{2-a+b} & \varepsilon^b \\
    \pm\varepsilon^{5+a-b} & \pm\varepsilon^2 & \pm\varepsilon^a \\
    \pm\varepsilon^{5-b} & \pm\varepsilon^{2-a} & \pm1 \\
  \end{pmatrix}, \quad
  M_d \propto
  \begin{pmatrix}
    \varepsilon^4 & \varepsilon^{2-a+b} & \varepsilon^b \\
    \pm\varepsilon^{4+a-b} & \pm\varepsilon^2 & \pm\varepsilon^a \\
    \pm\varepsilon^{4-b} & \pm\varepsilon^{2-a} & \pm1 \\
  \end{pmatrix}, \label{eq:fourthmassmtx}
\end{align}
where $\pm$ corresponds any possible combinations of signs and $a,b \in \{0,1,2,3,4,5\}$.
Note that it is always  possible to fix the signs of (1,1), (1,2) and (1,3) components to +1 by the basis transformation for right-handed quarks.
We set powers of $\varepsilon$ on diagonal components in up and down sector quark mass matrices to (5,3,0) and (3,2,0) for type I, (5,3,0) and (4,2,0) for type II, (5,2,0) and (3,2,0) for type III and (5,2,0) and (4,2,0) for type IV in order to realize their hierarchical masses.
Here, we use only six $\Gamma_6$ singlets, $\bm{1^0_0}$, $\bm{1^0_1}$, $\bm{1^0_2}$, $\bm{1^1_0}$, $\bm{1^1_1}$ and $\bm{1^1_2}$, as irreducible representations to make our analysis simple.
Note that again powers of $\varepsilon$ in mass matrices are determined by $Z_6$ charges of entities of mass matrices.
Thus mass matrices of each type can be led by the following assignments,
\begin{align}
  &\textrm{Type~I}: \notag \\
  &Q = (\bm{1}^{b~\textrm{mod~2}}_{b~\textrm{mod~3}},\bm{1}^{a~\textrm{mod~2}}_{a~\textrm{mod~3}},\bm{1^0_0}), ~ u_R = (\bm{1}^{5-b~\textrm{mod~2}}_{5-b~\textrm{mod~3}},\bm{1}^{3-a~\textrm{mod~2}}_{3-a~\textrm{mod~3}},\bm{1^0_0}), ~ d_R = (\bm{1}^{3-b~\textrm{mod~2}}_{3-b~\textrm{mod~3}},\bm{1}^{2-a~\textrm{mod~2}}_{2-a~\textrm{mod~3}},\bm{1^0_0}), \\
  &\textrm{Type~II}: \notag \\
  &Q = (\bm{1}^{b~\textrm{mod~2}}_{b~\textrm{mod~3}},\bm{1}^{a~\textrm{mod~2}}_{a~\textrm{mod~3}},\bm{1^0_0}), ~ u_R = (\bm{1}^{5-b~\textrm{mod~2}}_{5-b~\textrm{mod~3}},\bm{1}^{3-a~\textrm{mod~2}}_{3-a~\textrm{mod~3}},\bm{1^0_0}), ~ d_R = (\bm{1}^{4-b~\textrm{mod~2}}_{4-b~\textrm{mod~3}},\bm{1}^{2-a~\textrm{mod~2}}_{2-a~\textrm{mod~3}},\bm{1^0_0}), \\
  &\textrm{Type~III}: \notag \\
  &Q = (\bm{1}^{b~\textrm{mod~2}}_{b~\textrm{mod~3}},\bm{1}^{a~\textrm{mod~2}}_{a~\textrm{mod~3}},\bm{1^0_0}), ~ u_R = (\bm{1}^{5-b~\textrm{mod~2}}_{5-b~\textrm{mod~3}},\bm{1}^{2-a~\textrm{mod~2}}_{2-a~\textrm{mod~3}},\bm{1^0_0}), ~ d_R = (\bm{1}^{3-b~\textrm{mod~2}}_{3-b~\textrm{mod~3}},\bm{1}^{2-a~\textrm{mod~2}}_{2-a~\textrm{mod~3}},\bm{1^0_0}), \\
  &\textrm{Type~IV}: \notag \\
  &Q = (\bm{1}^{b~\textrm{mod~2}}_{b~\textrm{mod~3}},\bm{1}^{a~\textrm{mod~2}}_{a~\textrm{mod~3}},\bm{1^0_0}), ~ u_R = (\bm{1}^{5-b~\textrm{mod~2}}_{5-b~\textrm{mod~3}},\bm{1}^{2-a~\textrm{mod~2}}_{2-a~\textrm{mod~3}},\bm{1^0_0}), ~ d_R = (\bm{1}^{4-b~\textrm{mod~2}}_{4-b~\textrm{mod~3}},\bm{1}^{2-a~\textrm{mod~2}}_{2-a~\textrm{mod~3}},\bm{1^0_0}).
\end{align}
On the other hand, it is not always true that mass matrices in four types are definitely realized by the above assignments.
It depends on weights of the Yukawa couplings.
All of the singlet modular forms of $\Gamma_6$ with certain $Z_6$ charges do not exist for weights less than 14 as shown in appendix \ref{appendix:B}.
For instance, the modular forms of weight 12 belong to $\bm{1^1_1}$ do not exist.
Yukawa couplings of the weights less than 14 can lead to mass matrices with some zeros due to this shortage of the modular forms on low weights.
We study the case of Yukawa couplings of the weight 14 in subsection \ref{subsec:weight14} and one of the weights less than 14 in subsection \ref{subsec:weight<14}.


\subsection{Weight 14}
\label{subsec:weight14}

First of all, we study the models with Yukawa couplings of weight 14 to avoid  zero textures in mass matrices of four types.
We choose $\tau=3.2i$ as a benchmark point of the modulus.
At weight 14, seven singlet modular forms, $Y_{\bm{1^0_0}}^{(14)}$, $Y_{\bm{1^1_2i}}^{(14)}$, $Y_{\bm{1^0_1}}^{(14)}$, $Y_{\bm{1^1_0}}^{(14)}$, $Y_{\bm{1^0_2}}^{(14)}$, $Y_{\bm{1^1_1}}^{(14)}$ and $Y_{\bm{1^1_2ii}}^{(14)}$, exist and they are approximated by  $\varepsilon$ as
\begin{align}
  &Y_{\bm{1^0_0}}^{(14)}/Y_{\bm{1^0_0}}^{(14)} = 1 \rightarrow 1, \quad
  Y_{\bm{1^1_2i}}^{(14)}/Y_{\bm{1^0_0}}^{(14)} = 0.172 \rightarrow \varepsilon, \\
  &Y_{\bm{1^0_1}}^{(14)}/Y_{\bm{1^0_0}}^{(14)} = 0.0208 \rightarrow \varepsilon^2, \quad
  Y_{\bm{1^1_0}}^{(14)}/Y_{\bm{1^0_0}}^{(14)} = 0.00358 \rightarrow \varepsilon^3, \\
  &Y_{\bm{1^0_2}}^{(14)}/Y_{\bm{1^0_0}}^{(14)} = 0.000435 \rightarrow \varepsilon^4, \quad
  Y_{\bm{1^1_1}}^{(14)}/Y_{\bm{1^0_0}}^{(14)} = 0.0000746 \rightarrow \varepsilon^5, \\
  &Y_{\bm{1^1_2ii}}^{(14)}/Y_{\bm{1^0_0}}^{(14)} = 0.00000156 \rightarrow \varepsilon^7,
\end{align}
at $\tau=3.2i$.
Note that $Y_{\bm{1^1_2ii}}^{(14)} \sim \varepsilon^7$ originates from $Y_{\bm{1^1_0}}^{(6)} Y_{\bm{1^0_2}}^{(8)} \sim \varepsilon^3\cdot \varepsilon^4$ while $Y_{\bm{1^1_2i}}^{(14)} \sim \varepsilon$ originates from $Y_{\bm{1^1_2}}^{(6)} Y_{\bm{1^0_0}}^{(8)} \sim \varepsilon \cdot 1$.
$\varepsilon^n$ for $n>5$ can appear when the different modular forms of the same irreducible representations exist.
In what follows, we ignore $Y_{\bm{1^1_2ii}}^{(14)}$ because it belongs to the same representation as $Y_{\bm{1^1_2i}}^{(14)}$ and $Y_{\bm{1^1_2i}}^{(14)}>>Y_{\bm{1^1_2ii}}^{(14)}$.


\subsubsection{Type I: (5,3,0) and (3,2,0)}

The mass matrices of type I are given by
\begin{align}
  &M_u =
  \begin{pmatrix}
    \alpha^{11} Y_{\bm{1^1_1}}^{(14)} & \alpha^{12} Y_{\bm{1}^{3+a-b~\textrm{mod~2}}_{3+a-b~\textrm{mod~3}}}^{(14)} & \alpha^{13} Y_{\bm{1}^{6-b~\textrm{mod~2}}_{6-b~\textrm{mod~3}}}^{(14)} \\
    \alpha^{21} Y_{\bm{1}^{1-a+b~\textrm{mod~2}}_{1-a+b~\textrm{mod~3}}}^{(14)} & \alpha^{22} Y_{\bm{1^1_0}}^{(14)} & \alpha^{23} Y_{\bm{1}^{6-a~\textrm{mod~2}}_{6-a~\textrm{mod~3}}}^{(14)} \\
    \alpha^{31} Y_{\bm{1}^{1+b~\textrm{mod~2}}_{1+b~\textrm{mod~3}}}^{(14)} & \alpha^{32} Y_{\bm{1}^{3+a~\textrm{mod~2}}_{3+a~\textrm{mod~3}}}^{(14)} & \alpha^{33} Y_{\bm{1^0_0}}^{(14)} \\
  \end{pmatrix}, \label{eq:M_u530} \\
  &M_d =
  \begin{pmatrix}
    \beta^{11} Y_{\bm{1^1_0}}^{(14)} & \beta^{12} Y_{\bm{1}^{4+a-b~\textrm{mod~2}}_{4+a-b~\textrm{mod~3}}}^{(14)} & \beta^{13} Y_{\bm{1}^{6-b~\textrm{mod~2}}_{6-b~\textrm{mod~3}}}^{(14)} \\
    \beta^{21} Y_{\bm{1}^{3-a+b~\textrm{mod~2}}_{3-a+b~\textrm{mod~3}}}^{(14)} & \beta^{22} Y_{\bm{1^0_1}}^{(14)} & \beta^{23} Y_{\bm{1}^{6-a~\textrm{mod~2}}_{6-a~\textrm{mod~3}}}^{(14)} \\
    \beta^{31} Y_{\bm{1}^{3+b~\textrm{mod~2}}_{3+b~\textrm{mod~3}}}^{(14)} & \beta^{32} Y_{\bm{1}^{4+a~\textrm{mod~2}}_{4+a~\textrm{mod~3}}}^{(14)} & \beta^{33} Y_{\bm{1^0_0}}^{(14)} \\
  \end{pmatrix}, \label{eq:M_d320}
\end{align}
where $\alpha^{ij}$ and $\beta^{ij}$ are coupling constants which we restrict to $\pm1$.
The hierarchical mass matrices in Eq.~(\ref{eq:firstmassmtx}) can be obtained by choosing $+1$ or $-1$ appropriately in $\alpha^{ij}$ and $\beta^{ij}$.
As a result, we find best-fit mass matrices at $\tau=3.2i$,
\begin{align}
M_u/Y_{\bm{1^0_0}}^{(14)} =
\begin{pmatrix}
Y_{\bm{1^1_1}}^{(14)} & Y_{\bm{1^0_2}}^{(14)} & Y_{\bm{1^1_0}}^{(14)} \\
Y_{\bm{1^0_2}}^{(14)} & Y_{\bm{1^1_0}}^{(14)} & Y_{\bm{1^0_1}}^{(14)} \\
-Y_{\bm{1^0_1}}^{(14)} & -Y_{\bm{1^1_2i}}^{(14)} & Y_{\bm{1^0_0}}^{(14)} \\
\end{pmatrix}/Y_{\bm{1^0_0}}^{(14)}
&=
\begin{pmatrix}
0.0000746 & 0.000435 & 0.00358 \\
0.000435 & 0.00358 & 0.0208 \\
-0.0208 & -0.172 & 1 \\
\end{pmatrix} \notag \\
&\sim
\begin{pmatrix}
\varepsilon^5 & \varepsilon^4 & \varepsilon^3 \\
\varepsilon^4 & \varepsilon^3 & \varepsilon^2 \\
-\varepsilon^2 & -\varepsilon & 1 \\
\end{pmatrix}, \\
M_d/Y_{\bm{1^0_0}}^{(14)} =
\begin{pmatrix}
Y_{\bm{1^1_0}}^{(14)} & Y_{\bm{1^1_0}}^{(14)} & Y_{\bm{1^1_0}}^{(14)} \\
Y_{\bm{1^0_1}}^{(14)} & -Y_{\bm{1^0_1}}^{(14)} & -Y_{\bm{1^0_1}}^{(14)} \\
-Y_{\bm{1^0_0}}^{(14)} & Y_{\bm{1^0_0}}^{(14)} & -Y_{\bm{1^0_0}}^{(14)} \\
\end{pmatrix}/Y_{\bm{1^0_0}}^{(14)}
&=
\begin{pmatrix}
0.00358 & 0.00358 & 0.00358 \\
0.0208 & -0.0208 & -0.0208 \\
-1 & 1 & -1 \\
\end{pmatrix} \notag \\
&\sim
\begin{pmatrix}
\varepsilon^3 & \varepsilon^3 & \varepsilon^3 \\
\varepsilon^2 & -\varepsilon^2 & -\varepsilon^2 \\
-1 & 1 & -1 \\
\end{pmatrix}.
\end{align}
These mass matrices correspond to $a = 2$, $b = 3$ and can be realized by
\begin{align}
Q = (\bm{1^1_0},\bm{1^0_2},\bm{1^0_0}), \quad u_R = (\bm{1^0_2},\bm{1^1_1},\bm{1^0_0}), \quad d_R = (\bm{1^0_0},\bm{1^0_0},\bm{1^0_0}), \label{eq:Assign_typeI}
\end{align}
and their mass matrices are written by,
\begin{align}
M_u =
\begin{pmatrix}
\alpha^{11}Y_{\bm{1^1_1}}^{(14)} & \alpha^{12}Y_{\bm{1^0_2}}^{(14)} & \alpha^{13}Y_{\bm{1^1_0}}^{(14)} \\
\alpha^{21}Y_{\bm{1^0_2}}^{(14)} & \alpha^{22}Y_{\bm{1^1_0}}^{(14)} & \alpha^{23}Y_{\bm{1^0_1}}^{(14)} \\
\alpha^{31}Y_{\bm{1^0_1}}^{(14)} & \alpha^{32}Y_{\bm{1^1_2i}}^{(14)} & \alpha^{33}Y_{\bm{1^0_0}}^{(14)} \\
\end{pmatrix}, \quad
M_d =
\begin{pmatrix}
\beta^{11}Y_{\bm{1^1_0}}^{(14)} & \beta^{12}Y_{\bm{1^1_0}}^{(14)} & \beta^{13}Y_{\bm{1^1_0}}^{(14)} \\
\beta^{21}Y_{\bm{1^0_1}}^{(14)} & \beta^{22}Y_{\bm{1^0_1}}^{(14)} & \beta^{23}Y_{\bm{1^0_1}}^{(14)} \\
\beta^{31}Y_{\bm{1^0_0}}^{(14)} & \beta^{32}Y_{\bm{1^0_0}}^{(14)} & \beta^{33}Y_{\bm{1^0_0}}^{(14)} \\
\end{pmatrix},
\end{align}
with the following choises of $+1$ or $-1$ in coupling constants,
\begin{align}
\begin{pmatrix}
\alpha^{11} & \alpha^{12} & \alpha^{13} \\
\alpha^{21} & \alpha^{22} & \alpha^{23} \\
\alpha^{31} & \alpha^{32} & \alpha^{33} \\
\end{pmatrix} =
\begin{pmatrix}
1 & 1 & 1 \\
1 & 1 & 1 \\
-1 & -1 & 1 \\
\end{pmatrix}, \quad
\begin{pmatrix}
\beta^{11} & \beta^{12} & \beta^{13} \\
\beta^{21} & \beta^{22} & \beta^{23} \\
\beta^{31} & \beta^{32} & \beta^{33} \\
\end{pmatrix} =
\begin{pmatrix}
1 & 1 & 1 \\
1 & -1 & -1 \\
-1 & 1 & -1 \\
\end{pmatrix}. \label{eq:Choise_typeI}
\end{align}
They lead to the following quark mass ratios,
\begin{align}
&(m_u,m_c,m_t)/m_t = (2.11\times 10^{-5}, 7.07\times 10^{-3},1), \\
&(m_d,m_s,m_b)/m_b = (2.91\times 10^{-3}, 1.97\times 10^{-2},1),
\end{align}
and the absolute values of the CKM matrix elements,
\begin{align}
|V_{\textrm{CKM}}| =
\begin{pmatrix}
0.973 & 0.231 & 0.000681 \\
0.231 & 0.973 & 0.0270 \\
0.00690 & 0.0261 & 1.00 \\
\end{pmatrix}.
\end{align}

Results are shown in Table \ref{tab:type-I}.
Our purpose is to derive quark masses and mixing angles without fine-tuning.
Thus, we have fixed $\alpha^{ij},\beta^{ij}=\pm 1$ to make our point clear.
If we vary $\alpha^{ij},\beta^{ij}={\cal O}(1)$ without fixing $\alpha^{ij},\beta^{ij}=\pm 1$,  
we can obtain more realistic values.
Of course, we have ambiguity in normalization of modular forms, although 
we expect naturally that normalization factors would not lead a large hierarchy.
Our values appear at a high energy scale such as the GUT scale.
Renormalization group effects change values by some factors, although such radiative corrections 
may be realized by varying $\alpha^{ij},\beta^{ij}={\cal O}(1)$.

\begin{table}[H]
  \begin{center}
    \renewcommand{\arraystretch}{1.3}
    $\begin{array}{c|c|c} \hline
      & {\rm Obtained\ values} & {\rm Observed\ values} \\ \hline
      (m_u,m_c,m_t)/m_t & (2.11\times 10^{-5}, 7.07\times 10^{-3},1) & (1.26\times 10^{-5},7.38\times 10^{-3},1) \\ \hline
      (m_d,m_s,m_b)/m_b & (2.91\times 10^{-3}, 1.97\times 10^{-2},1) & (1.12\times 10^{-3},2.22\times 10^{-2},1) \\ \hline
      |V_{\rm CKM}| 
      &
      \begin{pmatrix}
0.973 & 0.231 & 0.000681 \\
0.231 & 0.973 & 0.0270 \\
0.00690 & 0.0261 & 1.00 \\
      \end{pmatrix}
      & 
      \begin{pmatrix}
        0.974 & 0.227 & 0.00361 \\
        0.226 & 0.973 & 0.0405 \\
        0.00854 & 0.0398 & 0.999 
      \end{pmatrix}\\ \hline
    \end{array}$
  \end{center}
  \caption{The mass ratios of the quarks and the absolute values of the CKM matrix elements at the benchmark point $\tau=3.2i$ in the best-fit model by Eqs.~(\ref{eq:Assign_typeI}) and (\ref{eq:Choise_typeI}) of type I with Yukawa couplings of weight 14.
  Observed values are shown in Ref.~\cite{Zyla:2020zbs}.}
\label{tab:type-I}
\end{table}


\subsubsection{Type II: (5,3,0) and (4,2,0)}

The mass matrices of type II are given by Eq.~(\ref{eq:M_u530}) and
\begin{align}
  M_d =
  \begin{pmatrix}
    \beta^{11} Y_{\bm{1^0_2}}^{(14)} & \beta^{12} Y_{\bm{1}^{4+a-b~\textrm{mod~2}}_{4+a-b~\textrm{mod~3}}}^{(14)} & \beta^{13} Y_{\bm{1}^{6-b~\textrm{mod~2}}_{6-b~\textrm{mod~3}}}^{(14)} \\
    \beta^{21} Y_{\bm{1}^{2-a+b~\textrm{mod~2}}_{2-a+b~\textrm{mod~3}}}^{(14)} & \beta^{22} Y_{\bm{1^0_1}}^{(14)} & \beta^{23} Y_{\bm{1}^{6-a~\textrm{mod~2}}_{6-a~\textrm{mod~3}}}^{(14)} \\
    \beta^{31} Y_{\bm{1}^{2+b~\textrm{mod~2}}_{2+b~\textrm{mod~3}}}^{(14)} & \beta^{32} Y_{\bm{1}^{4+a~\textrm{mod~2}}_{4+a~\textrm{mod~3}}}^{(14)} & \beta^{33} Y_{\bm{1^0_0}}^{(14)} \\
  \end{pmatrix}. \label{eq:M_d420}
\end{align}
The hierarchical mass matrices in Eq.~(\ref{eq:secondmassmtx}) can be obtained by choosing $+1$ or $-1$ appropriately in $\alpha^{ij}$ and $\beta^{ij}$.
As a result, we find best-fit mass matrices at $\tau=3.2i$,
\begin{align}
M_u/Y_{\bm{1^0_0}}^{(14)} =
\begin{pmatrix}
Y_{\bm{1^1_1}}^{(14)} & Y_{\bm{1^0_2}}^{(14)} & Y_{\bm{1^1_0}}^{(14)} \\
Y_{\bm{1^0_2}}^{(14)} & Y_{\bm{1^1_0}}^{(14)} & -Y_{\bm{1^0_1}}^{(14)} \\
-Y_{\bm{1^0_1}}^{(14)} & -Y_{\bm{1^1_2i}}^{(14)} & -Y_{\bm{1^0_0}}^{(14)} \\
\end{pmatrix}/Y_{\bm{1^0_0}}^{(14)}
&=
\begin{pmatrix}
0.0000746 & 0.000435 & 0.00358 \\
0.000435 & 0.00358 & -0.0208 \\
-0.0208 & -0.172 & -1 \\
\end{pmatrix} \notag \\
&\sim
\begin{pmatrix}
\varepsilon^5 & \varepsilon^4 & \varepsilon^3 \\
\varepsilon^4 & \varepsilon^3 & -\varepsilon^2 \\
-\varepsilon^2 & -\varepsilon & -1 \\
\end{pmatrix}, \\
M_d/Y_{\bm{1^0_0}}^{(14)} =
\begin{pmatrix}
Y_{\bm{1^0_2}}^{(14)} & Y_{\bm{1^1_0}}^{(14)} & Y_{\bm{1^1_0}}^{(14)} \\
-Y_{\bm{1^1_0}}^{(14)} & Y_{\bm{1^0_1}}^{(14)} & Y_{\bm{1^0_1}}^{(14)} \\
Y_{\bm{1^1_2i}}^{(14)} & Y_{\bm{1^0_0}}^{(14)} & -Y_{\bm{1^0_0}}^{(14)} \\
\end{pmatrix}/Y_{\bm{1^0_0}}^{(14)}
&=
\begin{pmatrix}
0.000435 & 0.00358 & 0.00358 \\
-0.00358 & 0.0208 & 0.0208 \\
0.172 & 1 & -1 \\
\end{pmatrix} \notag \\
&\sim
\begin{pmatrix}
\varepsilon^4 & \varepsilon^3 & \varepsilon^3 \\
-\varepsilon^3 & \varepsilon^2 & \varepsilon^2 \\
\varepsilon & 1 & -1 \\
\end{pmatrix}.
\end{align}
These mass matrices correspond to $a = 2$, $b = 3$ and can be realized by
\begin{align}
Q = (\bm{1^1_0},\bm{1^0_2},\bm{1^0_0}), \quad u_R = (\bm{1^0_2},\bm{1^1_1},\bm{1^0_0}), \quad d_R = (\bm{1^1_1},\bm{1^0_0},\bm{1^0_0}), \label{eq:Assign_typeII}
\end{align}
and their mass matrices are written by,
\begin{align}
M_u =
\begin{pmatrix}
\alpha^{11}Y_{\bm{1^1_1}}^{(14)} & \alpha^{12}Y_{\bm{1^0_2}}^{(14)} & \alpha^{13}Y_{\bm{1^1_0}}^{(14)} \\
\alpha^{21}Y_{\bm{1^0_2}}^{(14)} & \alpha^{22}Y_{\bm{1^1_0}}^{(14)} & \alpha^{23}Y_{\bm{1^0_1}}^{(14)} \\
\alpha^{31}Y_{\bm{1^0_1}}^{(14)} & \alpha^{32}Y_{\bm{1^1_2i}}^{(14)} & \alpha^{33}Y_{\bm{1^0_0}}^{(14)} \\
\end{pmatrix}, \quad
M_d =
\begin{pmatrix}
\beta^{11}Y_{\bm{1^0_2}}^{(14)} & \beta^{12}Y_{\bm{1^1_0}}^{(14)} & \beta^{13}Y_{\bm{1^1_0}}^{(14)} \\
\beta^{21}Y_{\bm{1^1_0}}^{(14)} & \beta^{22}Y_{\bm{1^0_1}}^{(14)} & \beta^{23}Y_{\bm{1^0_1}}^{(14)} \\
\beta^{31}Y_{\bm{1^1_2i}}^{(14)} & \beta^{32}Y_{\bm{1^0_0}}^{(14)} & \beta^{33}Y_{\bm{1^0_0}}^{(14)} \\
\end{pmatrix},
\end{align}
with the following choises of $+1$ or $-1$ in coupling constants,
\begin{align}
\begin{pmatrix}
\alpha^{11} & \alpha^{12} & \alpha^{13} \\
\alpha^{21} & \alpha^{22} & \alpha^{23} \\
\alpha^{31} & \alpha^{32} & \alpha^{33} \\
\end{pmatrix} =
\begin{pmatrix}
1 & 1 & 1 \\
1 & 1 & -1 \\
-1 & -1 & -1 \\
\end{pmatrix}, \quad
\begin{pmatrix}
\beta^{11} & \beta^{12} & \beta^{13} \\
\beta^{21} & \beta^{22} & \beta^{23} \\
\beta^{31} & \beta^{32} & \beta^{33} \\
\end{pmatrix} =
\begin{pmatrix}
1 & 1 & 1 \\
-1 & 1 & 1 \\
1 & 1 & -1 \\
\end{pmatrix}. \label{eq:Choise_typeII}
\end{align}
They lead to the following quark mass ratios,
\begin{align}
&(m_u,m_c,m_t)/m_t = (2.14\times 10^{-5}, 7.00\times 10^{-3},1), \\
&(m_d,m_s,m_b)/m_b = (7.16\times 10^{-4}, 2.11\times 10^{-2},1),
\end{align}
and the absolute values of the CKM matrix elements,
\begin{align}
|V_{\textrm{CKM}}| =
\begin{pmatrix}
0.982 & 0.190 & 0.00309 \\
0.190 & 0.982 & 0.0200 \\
0.00683 & 0.0191 & 1.00 \\
\end{pmatrix}.
\end{align}
Results are shown in Table \ref{tab:type-II}.
\begin{table}[H]
  \begin{center}
    \renewcommand{\arraystretch}{1.3}
    $\begin{array}{c|c|c} \hline
      & {\rm Obtained\ values} & {\rm Observed\ values} \\ \hline
      (m_u,m_c,m_t)/m_t & (2.14\times 10^{-5}, 7.00\times 10^{-3},1) & (1.26\times 10^{-5},7.38\times 10^{-3},1) \\ \hline
      (m_d,m_s,m_b)/m_b & (7.16\times 10^{-4}, 2.11\times 10^{-2},1) & (1.12\times 10^{-3},2.22\times 10^{-2},1) \\ \hline
      |V_{\rm CKM}| 
      &
      \begin{pmatrix}
0.982 & 0.190 & 0.00309 \\
0.190 & 0.982 & 0.0200 \\
0.00683 & 0.0191 & 1.00 \\
      \end{pmatrix}
      & 
      \begin{pmatrix}
        0.974 & 0.227 & 0.00361 \\
        0.226 & 0.973 & 0.0405 \\
        0.00854 & 0.0398 & 0.999 
      \end{pmatrix}\\ \hline
    \end{array}$
  \end{center}
  \caption{The mass ratios of the quarks and the absolute values of the CKM matrix elements at the benchmark point $\tau=3.2i$ in the best-fit model by Eqs.~(\ref{eq:Assign_typeII}) and (\ref{eq:Choise_typeII}) of type II with Yukawa couplings of weight 14.
  Observed values are shown in Ref.~\cite{Zyla:2020zbs}.}
\label{tab:type-II}
\end{table}


\subsubsection{Type III: (5,2,0) and (3,2,0)}

The mass matrices of type III are given by Eq.~(\ref{eq:M_d320}) and
\begin{align}
  &M_u =
  \begin{pmatrix}
    \alpha^{11} Y_{\bm{1^1_1}}^{(14)} & \alpha^{12} Y_{\bm{1}^{4+a-b~\textrm{mod~2}}_{4+a-b~\textrm{mod~3}}}^{(14)} & \alpha^{13} Y_{\bm{1}^{6-b~\textrm{mod~2}}_{6-b~\textrm{mod~3}}}^{(14)} \\
    \alpha^{21} Y_{\bm{1}^{1-a+b~\textrm{mod~2}}_{1-a+b~\textrm{mod~3}}}^{(14)} & \alpha^{22} Y_{\bm{1^0_1}}^{(14)} & \alpha^{23} Y_{\bm{1}^{6-a~\textrm{mod~2}}_{6-a~\textrm{mod~3}}}^{(14)} \\
    \alpha^{31} Y_{\bm{1}^{1+b~\textrm{mod~2}}_{1+b~\textrm{mod~3}}}^{(14)} & \alpha^{32} Y_{\bm{1}^{4+a~\textrm{mod~2}}_{4+a~\textrm{mod~3}}}^{(14)} & \alpha^{33} Y_{\bm{1^0_0}}^{(14)} \\
  \end{pmatrix}. \label{eq:M_u520}
\end{align}
The hierarchical mass matrices in Eq.~(\ref{eq:thirdmassmtx}) can be obtained by choosing $+1$ or $-1$ appropriately in $\alpha^{ij}$ and $\beta^{ij}$.
As a result, we find best-fit mass matrices at $\tau=3.2i$,
\begin{align}
M_u/Y_{\bm{1^0_0}}^{(14)} =
\begin{pmatrix}
Y_{\bm{1^1_1}}^{(14)} & Y_{\bm{1^1_0}}^{(14)} & Y_{\bm{1^1_0}}^{(14)} \\
Y_{\bm{1^0_2}}^{(14)} & -Y_{\bm{1^0_1}}^{(14)} & -Y_{\bm{1^0_1}}^{(14)} \\
Y_{\bm{1^0_1}}^{(14)} & Y_{\bm{1^0_0}}^{(14)} & -Y_{\bm{1^0_0}}^{(14)} \\
\end{pmatrix}/Y_{\bm{1^0_0}}^{(14)}
&=
\begin{pmatrix}
0.0000746 & 0.00358 & 0.00358 \\
0.000435 & -0.0208 & -0.0208 \\
0.0208 & 1 & -1 \\
\end{pmatrix} \notag \\
&\sim
\begin{pmatrix}
\varepsilon^5 & \varepsilon^3 & \varepsilon^3 \\
\varepsilon^4 & -\varepsilon^2 & -\varepsilon^2 \\
\varepsilon^2 & 1 & -1 \\
\end{pmatrix}, \\
M_d/Y_{\bm{1^0_0}}^{(14)} =
\begin{pmatrix}
Y_{\bm{1^1_0}}^{(14)} & Y_{\bm{1^1_0}}^{(14)} & Y_{\bm{1^1_0}}^{(14)} \\
Y_{\bm{1^0_1}}^{(14)} & Y_{\bm{1^0_1}}^{(14)} & -Y_{\bm{1^0_1}}^{(14)} \\
Y_{\bm{1^0_0}}^{(14)} & -Y_{\bm{1^0_0}}^{(14)} & Y_{\bm{1^0_0}}^{(14)} \\
\end{pmatrix}/Y_{\bm{1^0_0}}^{(14)}
&=
\begin{pmatrix}
0.00358 & 0.00358 & 0.00358 \\
0.0208 & 0.0208 & -0.0208 \\
1 & -1 & 1 \\
\end{pmatrix} \notag \\
&\sim
\begin{pmatrix}
\varepsilon^3 & \varepsilon^3 & \varepsilon^3 \\
\varepsilon^2 & \varepsilon^2 & -\varepsilon^2 \\
1 & -1 & 1 \\
\end{pmatrix}.
\end{align}
These mass matrices correspond to $a = 2$, $b = 3$ and can be realized by
\begin{align}
Q = (\bm{1^1_0},\bm{1^0_2},\bm{1^0_0}), \quad u_R = (\bm{1^0_2},\bm{1^0_0},\bm{1^0_0}), \quad d_R = (\bm{1^0_0},\bm{1^0_0},\bm{1^0_0}), \label{eq:Assign_typeIII}
\end{align}
and their mass matrices are written by,
\begin{align}
M_u =
\begin{pmatrix}
\alpha^{11}Y_{\bm{1^1_1}}^{(14)} & \alpha^{12}Y_{\bm{1^1_0}}^{(14)} & \alpha^{13}Y_{\bm{1^1_0}}^{(14)} \\
\alpha^{21}Y_{\bm{1^0_2}}^{(14)} & \alpha^{22}Y_{\bm{1^0_1}}^{(14)} & \alpha^{23}Y_{\bm{1^0_1}}^{(14)} \\
\alpha^{31}Y_{\bm{1^0_1}}^{(14)} & \alpha^{32}Y_{\bm{1^0_0}}^{(14)} & \alpha^{33}Y_{\bm{1^0_0}}^{(14)} \\
\end{pmatrix}, \quad
M_d =
\begin{pmatrix}
\beta^{11}Y_{\bm{1^1_0}}^{(14)} & \beta^{12}Y_{\bm{1^1_0}}^{(14)} & \beta^{13}Y_{\bm{1^1_0}}^{(14)} \\
\beta^{21}Y_{\bm{1^0_1}}^{(14)} & \beta^{22}Y_{\bm{1^0_1}}^{(14)} & \beta^{23}Y_{\bm{1^0_1}}^{(14)} \\
\beta^{31}Y_{\bm{1^0_0}}^{(14)} & \beta^{32}Y_{\bm{1^0_0}}^{(14)} & \beta^{33}Y_{\bm{1^0_0}}^{(14)} \\
\end{pmatrix},
\end{align}
with the following choises of $+1$ or $-1$ in coupling constants,
\begin{align}
\begin{pmatrix}
\alpha^{11} & \alpha^{12} & \alpha^{13} \\
\alpha^{21} & \alpha^{22} & \alpha^{23} \\
\alpha^{31} & \alpha^{32} & \alpha^{33} \\
\end{pmatrix} =
\begin{pmatrix}
1 & 1 & 1 \\
1 & -1 & -1 \\
1 & 1 & -1 \\
\end{pmatrix}, \quad
\begin{pmatrix}
\beta^{11} & \beta^{12} & \beta^{13} \\
\beta^{21} & \beta^{22} & \beta^{23} \\
\beta^{31} & \beta^{32} & \beta^{33} \\
\end{pmatrix} =
\begin{pmatrix}
1 & 1 & 1 \\
1 & 1 & -1 \\
1 & -1 & 1 \\
\end{pmatrix}. \label{eq:Choise_typeIII}
\end{align}
They lead to the following quark mass ratios,
\begin{align}
&(m_u,m_c,m_t)/m_t = (1.04\times 10^{-4}, 2.12\times 10^{-2},1), \\
&(m_d,m_s,m_b)/m_b = (2.91\times 10^{-3}, 1.97\times 10^{-2},1),
\end{align}
and the absolute values of the CKM matrix elements,
\begin{align}
|V_{\textrm{CKM}}| =
\begin{pmatrix}
0.967 & 0.255 & 0.00000171 \\
0.255 & 0.967 & 0.00706 \\
0.00180 & 0.00682 & 1.00 \\
\end{pmatrix}.
\end{align}
Results are shown in Table \ref{tab:type-III}.
\begin{table}[H]
  \begin{center}
    \renewcommand{\arraystretch}{1.3}
    $\begin{array}{c|c|c} \hline
      & {\rm Obtained\ values} & {\rm Observed\ values} \\ \hline
      (m_u,m_c,m_t)/m_t & (1.04\times 10^{-4}, 2.12\times 10^{-2},1) & (1.26\times 10^{-5},7.38\times 10^{-3},1) \\ \hline
      (m_d,m_s,m_b)/m_b & (2.91\times 10^{-3}, 1.97\times 10^{-2},1) & (1.12\times 10^{-3},2.22\times 10^{-2},1) \\ \hline
      |V_{\rm CKM}| 
      &
      \begin{pmatrix}
0.967 & 0.255 & 0.00000171 \\
0.255 & 0.967 & 0.00706 \\
0.00180 & 0.00682 & 1.00 \\
      \end{pmatrix}
      & 
      \begin{pmatrix}
        0.974 & 0.227 & 0.00361 \\
        0.226 & 0.973 & 0.0405 \\
        0.00854 & 0.0398 & 0.999 
      \end{pmatrix}\\ \hline
    \end{array}$
  \end{center}
  \caption{The mass ratios of the quarks and the absolute values of the CKM matrix elements at the benchmark point $\tau=3.2i$ in the best-fit model by Eqs.~(\ref{eq:Assign_typeIII}) and (\ref{eq:Choise_typeIII}) of type III with Yukawa couplings of weight 14.
  Observed values are shown in Ref.~\cite{Zyla:2020zbs}.}
\label{tab:type-III}
\end{table}


\subsubsection{Type IV: (5,2,0) and (4,2,0)}

The mass matrices of type IV are given by Eqs.~(\ref{eq:M_u520}) and (\ref{eq:M_d420}).
The hierarchical mass matrices in Eq.~(\ref{eq:fourthmassmtx}) can be obtained by choosing $+1$ or $-1$ appropriately in $\alpha^{ij}$ and $\beta^{ij}$.
As a result, we find best-fit mass matrices at $\tau=3.2i$,
\begin{align}
M_u/Y_{\bm{1^0_0}}^{(14)} =
\begin{pmatrix}
Y_{\bm{1^1_1}}^{(14)} & Y_{\bm{1^1_0}}^{(14)} & Y_{\bm{1^0_0}}^{(14)} \\
Y_{\bm{1^0_2}}^{(14)} & Y_{\bm{1^0_1}}^{(14)} & Y_{\bm{1^1_1}}^{(14)} \\
Y_{\bm{1^1_1}}^{(14)} & -Y_{\bm{1^1_0}}^{(14)} & -Y_{\bm{1^0_0}}^{(14)} \\
\end{pmatrix}/Y_{\bm{1^0_0}}^{(14)}
&=
\begin{pmatrix}
0.0000746 & 0.00358 & 1 \\
0.000435 & 0.0208 & 0.0000746 \\
0.0000746 & -0.00358 & -1 \\
\end{pmatrix} \notag \\
&\sim
\begin{pmatrix}
\varepsilon^5 & \varepsilon^3 & 1 \\
\varepsilon^4 & \varepsilon^2 & \varepsilon^5 \\
\varepsilon^5 & -\varepsilon^3 & -1 \\
\end{pmatrix}, \\
M_d/Y_{\bm{1^0_0}}^{(14)} =
\begin{pmatrix}
Y_{\bm{1^0_2}}^{(14)} & Y_{\bm{1^1_0}}^{(14)} & Y_{\bm{1^0_0}}^{(14)} \\
Y_{\bm{1^1_0}}^{(14)} & -Y_{\bm{1^0_1}}^{(14)} & -Y_{\bm{1^1_1}}^{(14)} \\
Y_{\bm{1^0_2}}^{(14)} & Y_{\bm{1^1_0}}^{(14)} & -Y_{\bm{1^0_0}}^{(14)} \\
\end{pmatrix}/Y_{\bm{1^0_0}}^{(14)}
&=
\begin{pmatrix}
0.000435 & 0.00358 & 1 \\
0.00358 & -0.0208 & -0.0000746 \\
0.000435 & 0.00358 & -1 \\
\end{pmatrix} \notag \\
&\sim
\begin{pmatrix}
\varepsilon^4 & \varepsilon^3 & 1 \\
\varepsilon^3 & -\varepsilon^2 & -\varepsilon^5 \\
\varepsilon^4 & \varepsilon^3 & -1 \\
\end{pmatrix}.
\end{align}
These mass matrices correspond to $a = 5$, $b = 0$ and can be realized by
\begin{align}
Q = (\bm{1^0_0},\bm{1^1_2},\bm{1^0_0}), \quad u_R = (\bm{1^1_2},\bm{1^1_0},\bm{1^0_0}), \quad d_R = (\bm{1^0_1},\bm{1^1_0},\bm{1^0_0}), \label{eq:Assign_typeIV}
\end{align}
and their mass matrices are written by,
\begin{align}
M_u =
\begin{pmatrix}
\alpha^{11}Y_{\bm{1^1_1}}^{(14)} & \alpha^{12}Y_{\bm{1^1_0}}^{(14)} & \alpha^{13}Y_{\bm{1^0_0}}^{(14)} \\
\alpha^{21}Y_{\bm{1^0_2}}^{(14)} & \alpha^{22}Y_{\bm{1^0_1}}^{(14)} & \alpha^{23}Y_{\bm{1^1_1}}^{(14)} \\
\alpha^{31}Y_{\bm{1^1_1}}^{(14)} & \alpha^{32}Y_{\bm{1^1_0}}^{(14)} & \alpha^{33}Y_{\bm{1^0_0}}^{(14)} \\
\end{pmatrix}, \quad
M_d =
\begin{pmatrix}
\beta^{11}Y_{\bm{1^0_2}}^{(14)} & \beta^{12}Y_{\bm{1^1_0}}^{(14)} & \beta^{13}Y_{\bm{1^0_0}}^{(14)} \\
\beta^{21}Y_{\bm{1^1_0}}^{(14)} & \beta^{22}Y_{\bm{1^0_1}}^{(14)} & \beta^{23}Y_{\bm{1^1_1}}^{(14)} \\
\beta^{31}Y_{\bm{1^0_2}}^{(14)} & \beta^{32}Y_{\bm{1^1_0}}^{(14)} & \beta^{33}Y_{\bm{1^0_0}}^{(14)} \\
\end{pmatrix},
\end{align}
with the following choises of $+1$ or $-1$ in coupling constants,
\begin{align}
\begin{pmatrix}
\alpha^{11} & \alpha^{12} & \alpha^{13} \\
\alpha^{21} & \alpha^{22} & \alpha^{23} \\
\alpha^{31} & \alpha^{32} & \alpha^{33} \\
\end{pmatrix} =
\begin{pmatrix}
1 & 1 & 1 \\
1 & 1 & 1 \\
1 & -1 & -1 \\
\end{pmatrix}, \quad
\begin{pmatrix}
\beta^{11} & \beta^{12} & \beta^{13} \\
\beta^{21} & \beta^{22} & \beta^{23} \\
\beta^{31} & \beta^{32} & \beta^{33} \\
\end{pmatrix} =
\begin{pmatrix}
1 & 1 & 1 \\
1 & -1 & -1 \\
1 & 1 & -1 \\
\end{pmatrix}. \label{eq:Choise_typeIV}
\end{align}
They lead to the following quark mass ratios,
\begin{align}
&(m_u,m_c,m_t)/m_t = (7.46\times 10^{-5}, 1.47\times 10^{-2},1), \\
&(m_d,m_s,m_b)/m_b = (1.01\times 10^{-3}, 1.54\times 10^{-2},1),
\end{align}
and the absolute values of the CKM matrix elements,
\begin{align}
|V_{\textrm{CKM}}| =
\begin{pmatrix}
0.974 & 0.226 & 0.0000000194 \\
0.226 & 0.974 & 0.000158 \\
0.0000358 & 0.000154 & 1.00 \\
\end{pmatrix}.
\end{align}
Results are shown in Table \ref{tab:type-IV}.
\begin{table}[H]
  \begin{center}
    \renewcommand{\arraystretch}{1.3}
    $\begin{array}{c|c|c} \hline
      & {\rm Obtained\ values} & {\rm Observed\ values} \\ \hline
      (m_u,m_c,m_t)/m_t & (7.46\times 10^{-5}, 1.47\times 10^{-2},1) & (1.26\times 10^{-5},7.38\times 10^{-3},1) \\ \hline
      (m_d,m_s,m_b)/m_b & (1.01\times 10^{-3}, 1.54\times 10^{-2},1) & (1.12\times 10^{-3},2.22\times 10^{-2},1) \\ \hline
      |V_{\rm CKM}| 
      &
      \begin{pmatrix}
0.974 & 0.226 & 0.0000000194 \\
0.226 & 0.974 & 0.000158 \\
0.0000358 & 0.000154 & 1.00 \\
      \end{pmatrix}
      & 
      \begin{pmatrix}
        0.974 & 0.227 & 0.00361 \\
        0.226 & 0.973 & 0.0405 \\
        0.00854 & 0.0398 & 0.999 
      \end{pmatrix}\\ \hline
    \end{array}$
  \end{center}
  \caption{The mass ratios of the quarks and the absolute values of the CKM matrix elements at the benchmark point $\tau=3.2i$ in the best-fit model by Eqs.~(\ref{eq:Assign_typeIV}) and (\ref{eq:Choise_typeIV}) of type IV with Yukawa couplings of weight 14.
  Observed values are shown in Ref.~\cite{Zyla:2020zbs}.}
\label{tab:type-IV}
\end{table}


\subsection{Weights less than 14}
\label{subsec:weight<14}

Next we study the models with Yukawa couplings of weights less than 14.
In this case, some of entities in mass matrices vanish because there do not exist 
modular forms of proper weights and representations.
As an example, let us consider the case that Yukawa couplings for the up sector have the  weight 8 and ones for the down sector have the  weight 10.
We choose $\tau=3.7i$ as a benchmark point of the modulus.
At weight 8, four singlet modular forms, $Y_{\bm{1^0_0}}^{(8)}$, $Y_{\bm{1^1_2}}^{(8)}$, $Y_{\bm{1^0_1}}^{(8)}$ and $Y_{\bm{1^0_2}}^{(8)}$, exist and they are approximated by $\varepsilon$ as
\begin{align}
  &Y_{\bm{1^0_0}}^{(8)}/Y_{\bm{1^0_0}}^{(8)} = 1 \rightarrow 1, \quad
  Y_{\bm{1^1_2}}^{(8)}/Y_{\bm{1^0_0}}^{(8)} = -0.0719 \rightarrow \varepsilon, \\
  &Y_{\bm{1^0_1}}^{(8)}/Y_{\bm{1^0_0}}^{(8)} = 0.00732 \rightarrow \varepsilon^2, \quad
  Y_{\bm{1^0_2}}^{(8)}/Y_{\bm{1^0_0}}^{(8)} = 0.0000535 \rightarrow \varepsilon^4,
\end{align}
at $\tau=3.7i$.
At weight 10, five singlet modular forms, $Y_{\bm{1^0_0}}^{(10)}$, $Y_{\bm{1^1_2}}^{(10)}$, $Y_{\bm{1^0_1}}^{(10)}$, $Y_{\bm{1^1_0}}^{(10)}$ and $Y_{\bm{1^1_1}}^{(10)}$, exist and they are approximated by  $\varepsilon$ as
\begin{align}
  &Y_{\bm{1^0_0}}^{(10)}/Y_{\bm{1^0_0}}^{(10)} = 1 \rightarrow 1, \quad
  Y_{\bm{1^1_2}}^{(10)}/Y_{\bm{1^0_0}}^{(10)} = 0.102 \rightarrow \varepsilon, \\
  &Y_{\bm{1^0_1}}^{(10)}/Y_{\bm{1^0_0}}^{(10)} = 0.00732 \rightarrow \varepsilon^2, \quad
  Y_{\bm{1^1_0}}^{(10)}/Y_{\bm{1^0_0}}^{(10)} = 0.000744 \rightarrow \varepsilon^3, \\
  &Y_{\bm{1^1_1}}^{(10)}/Y_{\bm{1^0_0}}^{(10)} = 0.00000544 \rightarrow \varepsilon^5, 
\end{align}
at $\tau=3.7i$.
As a result, we find the following best-fit mass matrices of type III,
\begin{align}
M_u/Y_{\bm{1^0_0}}^{(8)} =
\begin{pmatrix}
0 & 0 & Y_{\bm{1^0_1}}^{(8)} \\
Y_{\bm{1^0_2}}^{(8)} & Y_{\bm{1^0_1}}^{(8)} & Y_{\bm{1^1_2}}^{(8)} \\
0 & -Y_{\bm{1^1_2}}^{(8)} & -Y_{\bm{1^0_0}}^{(8)} \\
\end{pmatrix}/Y_{\bm{1^0_0}}^{(8)}
&=
\begin{pmatrix}
0 & 0 & 0.00732 \\
0.0000535 & 0.00732 & -0.0719 \\
0 & 0.0719 & -1 \\
\end{pmatrix} \notag \\
&\sim
\begin{pmatrix}
0 & 0 & \varepsilon^2 \\
\varepsilon^4 & \varepsilon^2 & -\varepsilon \\
0 & \varepsilon & -1 \\
\end{pmatrix}, \label{eq:Upweight8} \\
M_d/Y_{\bm{1^0_0}}^{(10)} =
\begin{pmatrix}
Y_{\bm{1^1_0}}^{(10)} & Y_{\bm{1^1_0}}^{(10)} & Y_{\bm{1^0_1}}^{(10)} \\
-Y_{\bm{1^0_1}}^{(10)} & -Y_{\bm{1^0_1}}^{(10)} & Y_{\bm{1^1_2}}^{(10)} \\
Y_{\bm{1^1_2}}^{(10)} & -Y_{\bm{1^1_2}}^{(10)} & Y_{\bm{1^0_0}}^{(10)} \\
\end{pmatrix}/Y_{\bm{1^0_0}}^{(10)}
&=
\begin{pmatrix}
0.000744 & 0.000744 & 0.00732 \\
-0.00732 & -0.00732 & 0.102 \\
0.102 & -0.102 & 1 \\
\end{pmatrix} \notag \\
&\sim
\begin{pmatrix}
\varepsilon^3 & \varepsilon^3 & \varepsilon^2 \\
-\varepsilon^2 & -\varepsilon^2 & \varepsilon \\
\varepsilon & -\varepsilon & 1 \\
\end{pmatrix}.
\end{align}
These mass matrices correspond to $a = 1$, $b = 2$ and can be realized by
\begin{align}
Q = (\bm{1^0_2},\bm{1^1_1},\bm{1^0_0}), \quad u_R = (\bm{1^1_0},\bm{1^1_1},\bm{1^0_0}), \quad d_R = (\bm{1^1_1},\bm{1^1_1},\bm{1^0_0}), \label{eq:Assign_typeIII108}
\end{align}
and their mass matrices,
\begin{align}
M_u =
\begin{pmatrix}
0 & 0 & \alpha^{13}Y_{\bm{1^0_1}}^{(8)} \\
\alpha^{21}Y_{\bm{1^0_2}}^{(8)} & \alpha^{22}Y_{\bm{1^0_1}}^{(8)} & \alpha^{23}Y_{\bm{1^1_2}}^{(8)} \\
0 & \alpha^{32}Y_{\bm{1^1_2}}^{(8)} & \alpha^{33}Y_{\bm{1^0_0}}^{(8)} \\
\end{pmatrix}, \quad
M_d =
\begin{pmatrix}
\beta^{11}Y_{\bm{1^1_0}}^{(10)} & \beta^{12}Y_{\bm{1^1_0}}^{(10)} & \beta^{13}Y_{\bm{1^0_1}}^{(10)} \\
\beta^{21}Y_{\bm{1^0_1}}^{(10)} & \beta^{22}Y_{\bm{1^0_1}}^{(10)} & \beta^{23}Y_{\bm{1^1_2}}^{(10)} \\
\beta^{31}Y_{\bm{1^1_2}}^{(10)} & \beta^{32}Y_{\bm{1^1_2}}^{(10)} & \beta^{33}Y_{\bm{1^0_0}}^{(10)} \\
\end{pmatrix},
\end{align}
with the following choises of $+1$ or $-1$ in coupling constants,
\begin{align}
\begin{pmatrix}
\textrm{-} & \textrm{-} & \alpha^{13} \\
\alpha^{21} & \alpha^{22} & \alpha^{23} \\
\textrm{-} & \alpha^{32} & \alpha^{33} \\
\end{pmatrix} =
\begin{pmatrix}
\textrm{-} & \textrm{-} & 1 \\
1 & 1 & 1 \\
\textrm{-} & -1 & -1 \\
\end{pmatrix}, \quad
\begin{pmatrix}
\beta^{11} & \beta^{12} & \beta^{13} \\
\beta^{21} & \beta^{22} & \beta^{23} \\
\beta^{31} & \beta^{32} & \beta^{33} \\
\end{pmatrix} =
\begin{pmatrix}
1 & 1 & 1 \\
-1 & -1 & 1 \\
1 & -1 & 1 \\
\end{pmatrix}. \label{eq:Choise_typeIII108}
\end{align}
They lead to the following up quark and down quark mass ratios,
\begin{align}
&(m_u,m_c,m_t)/m_t = (1.27\times 10^{-5}, 2.18\times 10^{-3},1), \\
&(m_d,m_s,m_b)/m_b = (1.44\times 10^{-3}, 1.74\times 10^{-2},1),
\end{align}
and the absolute values of the CKM matrix elements,
\begin{align}
|V_{\textrm{CKM}}| =
\begin{pmatrix}
0.974 & 0.227 & 0.00741 \\
0.227 & 0.973 & 0.0300 \\
0.0140 & 0.0276 & 1.00 \\
\end{pmatrix}.
\end{align}
Results are shown in Table \ref{tab:small-weight}.
Thus it is also possible to realize a realistic quark flavor structure in the models with Yukawa couplings of weights less than 14 despite some zeros in mass matrices.
Here we studied the case that Yukawa couplings for the up sector have weight 8 and ones for the down sector have weight 10 but other cases may be available for realization of the quark flavor structure.
\begin{table}[H]
  \begin{center}
    \renewcommand{\arraystretch}{1.3}
    $\begin{array}{c|c|c} \hline
      & {\rm Obtained\ values} & {\rm Observed\ values} \\ \hline
      (m_u,m_c,m_t)/m_t & (1.27\times 10^{-5}, 2.18\times 10^{-3},1) & (1.26\times 10^{-5},7.38\times 10^{-3},1) \\ \hline
      (m_d,m_s,m_b)/m_b & (1.44\times 10^{-3}, 1.74\times 10^{-2},1) & (1.12\times 10^{-3},2.22\times 10^{-2},1) \\ \hline
      |V_{\rm CKM}| 
      &
      \begin{pmatrix}
0.974 & 0.227 & 0.00741 \\
0.227 & 0.973 & 0.0300 \\
0.0140 & 0.0276 & 1.00 \\
      \end{pmatrix}
      & 
      \begin{pmatrix}
        0.974 & 0.227 & 0.00361 \\
        0.226 & 0.973 & 0.0405 \\
        0.00854 & 0.0398 & 0.999 
      \end{pmatrix}\\ \hline
    \end{array}$
  \end{center}
  \caption{The mass ratios of the quarks and the absolute values of the CKM matrix elements at the benchmark point $\tau=3.7i$ in the best-fit model by Eqs.~(\ref{eq:Assign_typeIII108}) and (\ref{eq:Choise_typeIII108}) of type III with up sector Yukawa couplings of weight 8 and down sector Yukawa couplings of weight 10.
  Observed values are shown in Ref.~\cite{Zyla:2020zbs}.}
\label{tab:small-weight}
\end{table}


\subsection{Comment on the origin of $\Gamma_6$ modular symmetry}

Here we comment on a plausible origin of $\Gamma_6$ modular symmetry of the theories.
For example, some modular forms are derived from the torus compactification $T^2_1\times T^2_2\times T^2_3$ of the low-energy effective theory of the superstring theory with magnetic flux background \cite{Kobayashi:2018rad,Kobayashi:2018bff,Ohki:2020bpo,Kikuchi:2020frp,Kikuchi:2020nxn,
Kikuchi:2021ogn}.
The group $\Gamma_6$ may originate from one of $T^2_i$, while the others $T^2_j$ lead to a trivial symmetry.
Alternatively, 
since $\Gamma_6\simeq S_3\times A_4 \simeq \Gamma_2\times \Gamma_3$, it may be expected that $\Gamma_2\simeq S_3$ originates from one torus $T^2_1$ and $\Gamma_3\simeq A_4$ originates from another torus $T^2_2$ with the moduli stabilization $\tau_1=\tau_2 \equiv \tau$.
Then $T^2_3$ contributes to the group symmetry trivially.


\section{Conclusion}
\label{sec:4}

We have discussed the possibility to describe mass hierarchies of both up and down sector quarks as well as mixing angles without fine-tuning.
Describing the quark flavor structure requires $Z_n$ residual symmetry with $n\geq 6$.
We have studied the modular symmetric quark flavor models of $\Gamma_6\simeq S_3\times A_4$ in the vicinity of the cusp $\tau=i\infty$ where $Z_6$ residual symmetry remains.
Then the values of the modular forms become hierarchical as close to the cusp depending on their $Z_6$ residual charges.

In order to obtain viable models, we consider four types of quark mass matrices; the diagonal components in up and down sector quark mass matrices are written by $\varepsilon$ with the powers of $(5,3,0)$ and $(3,2,0)$ respectively for type I, $(5,3,0)$ and $(4,2,0)$ for type II, $(5,2,0)$ and $(3,2,0)$ for type III, and $(5,2,0)$ and $(4,2,0)$ for type IV.
The powers of non-diagonal components in up and down sector quark mass matrices have been treated as model depending values.
When we assign the irreducible representations into quarks and Higgs fields, powers of $\varepsilon$ in mass matrix components are  determined by residual charges of mass matrix components.
For simplicity, we have used only six singlets $\bm{1^0_0}$, $\bm{1^0_1}$, $\bm{1^0_2}$, $\bm{1^1_0}$, $\bm{1^1_1}$ and $\bm{1^1_2}$ as the irreducible representations of $\Gamma_6$.
In addition, we have restricted the values of the coupling constants to $\pm 1$ to avoid fine-tuning by them.

Firstly, we have investigated the case that up and down sector Yukawa couplings have weight 14.
In such cases, mass matrices have no zeros, that is, all of their components are written in terms of the modular forms for $\Gamma_6$ of weight 14.
Consequently, we have obtained viable models at $\tau=3.2i$ for each type without fine-tuning.

Second, we have shown the viable model in the case that Yukawa couplings of the up sector have weight 8 and ones of the down sector have weight 10.
In this case some components of mass matrices can become zero because there do not exist modular forms of proper weights and representations.
As a result, we have obtained the viable model at $\tau=3.7i$ despite three zeros in the up quark mass matrix, Eq.~(\ref{eq:Upweight8}).

Thus, the modular symmetric quark flavor models based on $\Gamma_6$ in the vicinity of the cusp $\tau=i\infty$ lead to successful quark mass matrices without fine-tuning.
As we have commented in the end of previous section, $\Gamma_6\simeq S_3\times A_4\simeq \Gamma_2\times \Gamma_3$ may originate from the torus compactification $T^2_1\times T^2_2\times T^2_3$ of the low-energy effective theory of superstring theory.
Motivated this point, the modular flavor models based on the direct product of  finite modular groups, $\Gamma_{N_1}\times\Gamma_{N_2}\times\Gamma_{N_3}$, may be interesting.
Also we can extend our analysis to the lepton sector.
We will study them in the near future.

In our models, the important parameter is the modulus $\tau$.
It must be stabilized such that the proper mass hierarchies are realized.
We would study such modulus stabilization elsewhere\footnote{
See for modulus stabilization in modular flavor symmetric models Refs.~\cite{Kobayashi:2019xvz,Ishiguro:2020tmo,Abe:2020vmv,Novichkov:2022wvg,Ishiguro:2022pde}
.}.

\vspace{1.5 cm}
\noindent
{\large\bf Acknowledgement}\\

This work was supported by JSPS KAKENHI Grant Numbers JP22J10172 (SK) and JP20J20388 (HU), 
and 
JST SPRING Grant Number JPMJSP2119(KN).

\appendix
\section*{Appendix}


\section{Tensor product of $\Gamma_6$ group}
\label{appendix:A}

Here, we give a review on group theoretical aspects of $\Gamma_6$.
The generators of $\Gamma_6$ are denoted by $S$ and $T$, and they satisfy the following algebraic relations:
\begin{align}
  S^2 = (ST)^3 = T^6 = ST^2ST^3ST^4ST^3 = 1, \quad S^2T = TS^2.
\end{align}
In $\Gamma_6$ group, there are 12 irreducible representations, six singlets $\bm{1^0_0}$, $\bm{1^0_1}$, $\bm{1^0_2}$, $\bm{1^1_0}$, $\bm{1^1_1}$ and $\bm{1^1_2}$, three doublets $\bm{2_0}$, $\bm{2_1}$ and $\bm{2_2}$, two triplets $\bm{3^0}$ and $\bm{3^1}$ and one six-dimensional representation $\bm{6}$.
Each irreducible representation is given by
\begin{align}
  &\bm{1^r_k}:~ S=(-1)^r, \quad T=(-1)^r\omega^k, \\
  &\bm{2_k}:~  S=\frac{1}{2}\begin{pmatrix}-1&\sqrt{3}\\\sqrt{3}& 1\\\end{pmatrix}, \quad T = \omega^k\begin{pmatrix}1&0\\0&-1\\\end{pmatrix}, \\
  &\bm{3^r}:~ (-1)^r\bm{a_3}, \quad (-1)^r\bm{b_3}, \\
  &\bm{6}:~\frac{1}{2}\begin{pmatrix}-\bm{a_3} & \sqrt{3}\bm{a_3} \\ \sqrt{3}\bm{a_3} & \bm{a_3}\\\end{pmatrix}, \quad T=\begin{pmatrix}\bm{b_3} & \bm{0} \\ \bm{0} & -\bm{b_3}\\\end{pmatrix},
\end{align}
where $r=0,1$, $k=0,1,2$ and
\begin{align}
  {\bm a}_3 = \frac{1}{3}
  \begin{pmatrix}
    -1 & 2 & 2 \\
    2 & -1 & 2 \\
    2 & 2 & -1 \\
  \end{pmatrix}, \quad
  {\bm b}_3 =
  \begin{pmatrix}
    1 & 0 & 0 \\
    0 & \omega & 0 \\
    0 & 0 & \omega^2 \\
  \end{pmatrix}.
\end{align}
In this basis, the Kronecker products between irreducible representations are:
\begin{align}
  &\bm{1^r_i} \otimes \bm{1^s_j} = \bm{1^t_m}, \quad
  \bm{1^r_i} \otimes \bm{2_j} = \bm{2_m}, \quad
  \bm{1^r_i} \otimes \bm{3^s} = \bm{3^t}, \quad
  \bm{1^r_i} \otimes \bm{6} = \bm{6}, \\
  &\bm{2_i} \otimes \bm{2_j} = \bm{1^0_m} \oplus \bm{1^1_m} \oplus \bm{2_m}, \quad
  \bm{2_i} \otimes \bm{3^r} = \bm{6}, \quad
  \bm{2_i} \otimes \bm{6} = \bm{3^0} \oplus \bm{3^1} \oplus \bm{6}, \\
  &\bm{3^r} \otimes \bm{3^s} = \bm{1^t_0} \oplus \bm{1^t_1} \oplus \bm{1^t_2} \oplus \bm{3^t_1} \oplus \bm{3^t_2}, \quad
  \bm{3^r} \otimes \bm{6} = \bm{2_0} \oplus \bm{2_1} \oplus \bm{2_2} \oplus \bm{6} \oplus \bm{6}, \\
  &\bm{6} \otimes \bm{6} = \bm{1^0_0} \oplus \bm{1^0_1} \oplus \bm{1^0_2} \oplus \bm{1^1_0} \oplus \bm{1^1_1} \oplus \bm{1^1_2} \oplus \bm{2_0} \oplus \bm{2_1} \oplus \bm{2_2} \oplus \bm{3^0} \oplus \bm{3^0} \oplus \bm{3^1} \oplus \bm{3^1} \oplus \bm{6} \oplus \bm{6},
\end{align}
where $i,j=0,1,2$, $r,s=0,1$, $m = i+j~(\textrm{mod}~3)$ and $t = r+s~(\textrm{mod}~2)$.
In the following, we show the Clebsch-Gordon (CG) coefficients of these products.
\begin{align}
  &(\alpha_1)_{\bm{1^r_i}} \otimes \begin{pmatrix}\beta^1\\\beta_2\\\end{pmatrix}_{\bm{2_j}} = \alpha_1P^r_2\begin{pmatrix}\beta_1\\\beta_2\\\end{pmatrix}_{\bm{2}_m}, \quad
  (\alpha_1)_{\bm{1^r_i}} \otimes \begin{pmatrix}\beta_1\\\beta_2\\\beta_3\\\end{pmatrix}_{\bm{3^s}} = \alpha_1 P^i_3\begin{pmatrix}\beta_1\\\beta_2\\\beta_3\\\end{pmatrix}_{\bm{3^t}}, \notag \\
  &(\alpha_1)_{\bm{1^r_i}} \otimes \begin{pmatrix}\beta_1\\ \beta_2\\ \beta_3\\ \beta_4\\ \beta_5\\ \beta_6\\\end{pmatrix}_{\bm{6}} = \alpha_1 P_6(r,i)\begin{pmatrix}\beta_1\\ \beta_2\\ \beta_3\\ \beta_4\\ \beta_5\\ \beta_6\\\end{pmatrix}_{\bm{6}}, \notag 
\end{align}
\begin{align}
  &\begin{pmatrix}\alpha_1\\\alpha_2\\\end{pmatrix}_{\bm{2_i}} \otimes \begin{pmatrix}\beta_1\\\beta_2\\\end{pmatrix}_{\bm{2_j}} = \frac{1}{\sqrt{2}}(\alpha_1\beta_1+\alpha_2\beta_2)_{\bm{1^0_m}} \oplus \frac{1}{\sqrt{2}}(\alpha_1\beta_2-\alpha_2\beta_1)_{\bm{1^1_m}} \oplus \frac{1}{\sqrt{2}}\begin{pmatrix}\alpha_1\beta_1-\alpha_2\beta_2\\-\alpha_1\beta_2-\alpha_2\beta_1\\\end{pmatrix}_{\bm{2_m}}, \notag \\
  &\begin{pmatrix}\alpha_1\\\alpha_2\\\end{pmatrix}_{\bm{2_i}} \otimes \begin{pmatrix}\beta_1\\\beta_2\\\beta_3\\\end{pmatrix}_{\bm{3^r}} = P_6(r,i)\begin{pmatrix}\alpha_1\beta_1\\ \alpha_1\beta_2\\ \alpha_1\beta_3\\ \alpha_2\beta_1\\ \alpha_2\beta_2\\ \alpha_2\beta_3\\\end{pmatrix}_{\bm{6}}, \notag \\
  &\begin{pmatrix}\alpha_1\\\alpha_2\\\end{pmatrix}_{\bm{2_i}} \otimes \begin{pmatrix}\beta_1\\ \beta_2\\ \beta_3\\ \beta_4\\ \beta_5\\ \beta_6\\\end{pmatrix}_{\bm{6}} = 
  \frac{P^i_3}{\sqrt{2}}\begin{pmatrix}\alpha_1\beta_1+\alpha_2\beta_4\\\alpha_1\beta_2+\alpha_2\beta_5\\\alpha_1\beta_3+\alpha_2\beta_6\\\end{pmatrix}_{\bm{3^0}} \oplus \frac{P^i_3}{\sqrt{2}}\begin{pmatrix}\alpha_1\beta_4-\alpha_2\beta_1\\\alpha_1\beta_5-\alpha_2\beta_2\\\alpha_1\beta_6-\alpha_2\beta_3\\\end{pmatrix}_{\bm{3^1}} \oplus \frac{P_6(0,i)}{\sqrt{2}}\begin{pmatrix}\alpha_1\beta_1-\alpha_2\beta_4\\\alpha_1\beta_2-\alpha_2\beta_5\\\alpha_1\beta_3-\alpha_2\beta_6\\-\alpha_1\beta_4-\alpha_2\beta_1\\-\alpha_1\beta_5-\alpha_2\beta_2\\-\alpha_1\beta_6-\alpha_2\beta_3\\\end{pmatrix}_{\bm{6}}, \notag 
\end{align}
\begin{align}
  &\begin{pmatrix}\alpha_1\\\alpha_2\\\alpha_3\\\end{pmatrix}_{\bm{3^r}} \otimes \begin{pmatrix}\beta_1\\\beta_2\\\beta_3\\\end{pmatrix}_{\bm{3^s}} 
  = \frac{1}{\sqrt{3}}(\alpha_1\beta_1+\alpha_2\beta_3+\alpha_3\beta_2)_{\bm{1^t_0}} 
  \oplus \frac{1}{\sqrt{3}}(\alpha_1\beta_2+\alpha_2\beta_1+\alpha_3\beta_3)_{\bm{1^t_1}} \notag \\
  &\oplus \frac{1}{\sqrt{3}}(\alpha_1\beta_3+\alpha_2\beta_2+\alpha_3\beta_1)_{\bm{1^t_2}} 
  \oplus \frac{1}{\sqrt{3}}\begin{pmatrix}2\alpha_1\beta_1-\alpha_2\beta_3-\alpha_3\beta_2 \\-\alpha_1\beta_2-\alpha_2\beta_1+2\alpha_3\beta_3 \\-\alpha_1\beta_3+2\alpha_2\beta_2-\alpha_3\beta_1 \\\end{pmatrix}_{\bm{3^t_1}} 
  \oplus \frac{1}{\sqrt{2}}\begin{pmatrix}-\alpha_2\beta_3+\alpha_3\beta_2 \\-\alpha_1\beta_2+\alpha_2\beta_1 \\\alpha_1\beta_3-\alpha_3\beta_1 \\\end{pmatrix}_{\bm{3^t_2}}, \notag \\
  &\begin{pmatrix}\alpha_1\\\alpha_2\\\alpha_3\\\end{pmatrix}_{\bm{3^r}} \otimes \begin{pmatrix}\beta_1\\ \beta_2\\ \beta_3\\ \beta_4\\ \beta_5\\ \beta_6\\\end{pmatrix}_{\bm{6}} 
  = \frac{P_2^r}{\sqrt{3}} \begin{pmatrix} \alpha_1\beta_1+\alpha_2\beta_3+\alpha_3\beta_2 \\\alpha_1\beta_4+\alpha_2\beta_6+\alpha_3\beta_5 \\ \end{pmatrix}_{\bm{2_0}} 
  \oplus \frac{P_2^r}{\sqrt{3}} \begin{pmatrix} \alpha_1\beta_2+\alpha_2\beta_1+\alpha_3\beta_3 \\\alpha_1\beta_5+\alpha_2\beta_4+\alpha_3\beta_6 \\ \end{pmatrix}_{\bm{2_1}} \notag \\
  &\oplus \frac{P_2^r}{\sqrt{3}} \begin{pmatrix} \alpha_1\beta_3+\alpha_2\beta_2+\alpha_3\beta_1 \\\alpha_1\beta_6+\alpha_2\beta_5+\alpha_3\beta_4 \\ \end{pmatrix}_{\bm{2_2}} 
  \oplus \frac{P_6(r,0)}{\sqrt{2}} \begin{pmatrix}  
  \alpha_1 \beta_1 - \alpha_3 \beta_2 \\
-\alpha_2 \beta_1 + \alpha_3 \beta_3 \\
-\alpha_1 \beta_3 + \alpha_2 \beta_2 \\
\alpha_1 \beta_4 - \alpha_3 \beta_5 \\
-\alpha_2 \beta_4 + \alpha_3 \beta_6 \\
-\alpha_1 \beta_6 + \alpha_2 \beta_5 \\  
  \end{pmatrix}_{\bm{6}} 
  \oplus \frac{P_6(r,0)}{\sqrt{2}} \begin{pmatrix}  
  \alpha_2 \beta_3 - \alpha_3 \beta_2 \\
\alpha_1 \beta_2 - \alpha_2 \beta_1 \\
-\alpha_1 \beta_3 + \alpha_3 \beta_1 \\
\alpha_2 \beta_6 - \alpha_3 \beta_5 \\
\alpha_1 \beta_5 - \alpha_2 \beta_4 \\
-\alpha_1 \beta_6 + \alpha_3 \beta_4 \\
  \end{pmatrix}_{\bm{6}}, \notag 
\end{align}
\begin{align}
  \begin{pmatrix}\alpha_1\\ \alpha_2\\ \alpha_3\\ \alpha_4\\ \alpha_5\\ \alpha_6\\\end{pmatrix}_{\bm{6}} \otimes \begin{pmatrix}\beta_1\\ \beta_2\\ \beta_3\\ \beta_4\\ \beta_5\\ \beta_6\\\end{pmatrix}_{\bm{6}}
  =~& 
  \begin{matrix}
  \frac{1}{\sqrt{6}}(\alpha_1 \beta_1 + \alpha_2 \beta_3 + \alpha_3 \beta_2 + \alpha_4 \beta_4 + \alpha_5 \beta_6 + \alpha_6 \beta_5)_{\bm{1^0_0}}~~ \\
  \oplus \frac{1}{\sqrt{6}}(\alpha_1 \beta_2 + \alpha_2 \beta_1 + \alpha_3 \beta_3 + \alpha_4 \beta_5 + \alpha_5 \beta_4 + \alpha_6 \beta_6)_{\bm{1^0_1}} \\
  \oplus \frac{1}{\sqrt{6}}(\alpha_1 \beta_3 + \alpha_2 \beta_2 + \alpha_3 \beta_1 + \alpha_4 \beta_6 + \alpha_5 \beta_5 + \alpha_6 \beta_4)_{\bm{1^0_2}} \\
  \oplus \frac{1}{\sqrt{6}}(\alpha_1 \beta_4 + \alpha_2 \beta_6 + \alpha_3 \beta_5 - \alpha_4 \beta_1 - \alpha_5 \beta_3 - \alpha_6 \beta_2)_{\bm{1^1_0}} \\
  \oplus \frac{1}{\sqrt{6}}(\alpha_1 \beta_5 + \alpha_2 \beta_4 + \alpha_3 \beta_6 - \alpha_4 \beta_2 - \alpha_5 \beta_1 - \alpha_6 \beta_3)_{\bm{1^1_1}} \\
  \oplus \frac{1}{\sqrt{6}}(\alpha_1 \beta_6 + \alpha_2 \beta_5 + \alpha_3 \beta_4 - \alpha_4 \beta_3 - \alpha_5 \beta_2 - \alpha_6 \beta_1)_{\bm{1^1_2}} \\
  \end{matrix} \notag \\
  &\oplus \textstyle\frac{1}{\sqrt{6}}\begin{pmatrix}\alpha_1 \beta_1 + \alpha_2 \beta_3 + \alpha_3 \beta_2 - \alpha_4 \beta_4 - \alpha_5 \beta_6 - \alpha_6 \beta_5 \\ -(\alpha_1 \beta_4 + \alpha_2 \beta_6 + \alpha_3 \beta_5 + \alpha_4 \beta_1 + \alpha_5 \beta_3 + \alpha_6 \beta_2)\end{pmatrix}_{\bm{2_0}} \notag \\
  &\oplus \textstyle\frac{1}{\sqrt{6}}\begin{pmatrix}\alpha_1 \beta_2 + \alpha_2 \beta_1 + \alpha_3 \beta_3 - \alpha_4 \beta_5 - \alpha_5 \beta_4 - \alpha_6 \beta_6 \\ -(\alpha_1 \beta_5 + \alpha_2 \beta_4 + \alpha_3 \beta_6 + \alpha_4 \beta_2 + \alpha_5 \beta_1 + \alpha_6 \beta_3 )\end{pmatrix}_{\bm{2_1}} \notag \\
  &\oplus \textstyle\frac{1}{\sqrt{6}}\begin{pmatrix} \alpha_1 \beta_3 + \alpha_2 \beta_2 + \alpha_3 \beta_1 - \alpha_4 \beta_6 - \alpha_5 \beta_5 - \alpha_6 \beta_4 \\ -(\alpha_1 \beta_6 + \alpha_2 \beta_5 + \alpha_3 \beta_4 + \alpha_4 \beta_3 + \alpha_5 \beta_2 + \alpha_6 \beta_1 ) \\ \end{pmatrix}_{\bm{2_2}} \notag \\ 
  &\oplus \textstyle\frac{1}{2\sqrt{3}}\begin{pmatrix} 2\alpha_1 \beta_1 - \alpha_2 \beta_3 - \alpha_3 \beta_2 + 2\alpha_4 \beta_4 - \alpha_5 \beta_6 - \alpha_6 \beta_5 \\
2\alpha_3 \beta_3 - \alpha_1 \beta_2 - \alpha_2 \beta_1 + 2\alpha_6 \beta_6 - \alpha_4 \beta_5 - \alpha_5 \beta_4 \\
2\alpha_2 \beta_2 - \alpha_1 \beta_3 - \alpha_3 \beta_1 + 2\alpha_5 \beta_5 - \alpha_4 \beta_6 - \alpha_6 \beta_4 \\ \end{pmatrix}_{\bm{3^0}} \notag \\ 
  &\oplus \textstyle\frac{1}{2}\begin{pmatrix} \alpha_2 \beta_3 - \alpha_3 \beta_2 + \alpha_5 \beta_6 - \alpha_6 \beta_5 \\
\alpha_1 \beta_2 - \alpha_2 \beta_1 + \alpha_4 \beta_5 - \alpha_5 \beta_4 \\
-\alpha_1 \beta_3 + \alpha_3 \beta_1 - \alpha_4 \beta_6 + \alpha_6 \beta_4 \\ \end{pmatrix}_{\bm{3^0}} \notag \\ 
  &\oplus \textstyle\frac{1}{2}\begin{pmatrix} \alpha_2 \beta_6 - \alpha_3 \beta_5 - \alpha_5 \beta_3 + \alpha_6 \beta_2 \\
\alpha_1 \beta_5 - \alpha_2 \beta_4 - \alpha_4 \beta_2 + \alpha_5 \beta_1 \\
-\alpha_1 \beta_6 + \alpha_3 \beta_4 + \alpha_4 \beta_3 - \alpha_6 \beta_1 \\ \end{pmatrix}_{\bm{3^1}} \notag \\ 
  &\oplus \textstyle\frac{1}{2\sqrt{3}}\begin{pmatrix} 2\alpha_1 \beta_4 - \alpha_2 \beta_6 - \alpha_3 \beta_5 - 2\alpha_4 \beta_1 + \alpha_5 \beta_3 + \alpha_6 \beta_2 \\
-\alpha_1 \beta_5 - \alpha_2 \beta_4 + 2\alpha_3 \beta_6 + \alpha_4 \beta_2 + \alpha_5 \beta_1 - 2\alpha_6 \beta_3 \\
-\alpha_1 \beta_6 + 2\alpha_2 \beta_5 - \alpha_3 \beta_4 + \alpha_4 \beta_3 - 2\alpha_5 \beta_2 + \alpha_6 \beta_1 \\ \end{pmatrix}_{\bm{3^1}} \notag \\ 
  &\oplus \textstyle\frac{1}{2\sqrt{3}}\begin{pmatrix} 2\alpha_1 \beta_1 - \alpha_2 \beta_3 - \alpha_3 \beta_2 - 2\alpha_4 \beta_4 + \alpha_5 \beta_6 + \alpha_6 \beta_5 \\
-\alpha_1 \beta_2 - \alpha_2 \beta_1 + 2\alpha_3 \beta_3 + \alpha_4 \beta_5 + \alpha_5 \beta_4 - 2\alpha_6 \beta_6 \\
-\alpha_1 \beta_3 + 2\alpha_2 \beta_2 - \alpha_3 \beta_1 + \alpha_4 \beta_6 - 2\alpha_5 \beta_5 + \alpha_6 \beta_4 \\
-2\alpha_1 \beta_4 + \alpha_2 \beta_6 + \alpha_3 \beta_5 - 2\alpha_4 \beta_1 + \alpha_5 \beta_3 + \alpha_6 \beta_2 \\
\alpha_1 \beta_5 + \alpha_2 \beta_4 - 2\alpha_3 \beta_6 + \alpha_4 \beta_2 + \alpha_5 \beta_1 - 2\alpha_6 \beta_3 \\
\alpha_1 \beta_6 - 2\alpha_2 \beta_5 + \alpha_3 \beta_4 + \alpha_4 \beta_3 - 2\alpha_5 \beta_2 + \alpha_6 \beta_1 \\\end{pmatrix}_{\bm{6}} \notag \\
  &\oplus \textstyle\frac{1}{2}\begin{pmatrix} \alpha_2 \beta_3 - \alpha_3 \beta_2 - \alpha_5 \beta_6 + \alpha_6 \beta_5 \\
\alpha_1 \beta_2 - \alpha_2 \beta_1 - \alpha_4 \beta_5 + \alpha_5 \beta_4 \\
-\alpha_1 \beta_3 + \alpha_3 \beta_1 + \alpha_4 \beta_6 - \alpha_6 \beta_4 \\
-\alpha_2 \beta_6 + \alpha_3 \beta_5 - \alpha_5 \beta_3 + \alpha_6 \beta_2 \\
-\alpha_1 \beta_5 + \alpha_2 \beta_4 - \alpha_4 \beta_2 + \alpha_5 \beta_1 \\
\alpha_1 \beta_6 - \alpha_3 \beta_4 + \alpha_4 \beta_3 - \alpha_6 \beta_1 \\ \end{pmatrix}_{\bm{6}}. \notag
\end{align}
Here we have used the notations,
\begin{align}
  P_2 = 
  \begin{pmatrix}
    0 & 1 \\
    -1 & 0 \\
  \end{pmatrix}, \quad
  P_3 = 
  \begin{pmatrix}
    0 & 0 & 1 \\
    1 & 0 & 0 \\
    0 & 1 & 0 \\
  \end{pmatrix}, \quad
  P_6(r,i) =
  \begin{pmatrix}
    \bm{0_3} & \bm{1_3} \\
    -\bm{1_3} & \bm{0_3} \\
  \end{pmatrix}^r
  \begin{pmatrix}
    P_3 & \bm{0_3} \\
    \bm{0_3} & P_3 \\
  \end{pmatrix}^i.
\end{align}
Further details can be found in Ref.~\cite{Li:2021buv}.


\section{Modular forms of $\Gamma_6$}
\label{appendix:B}

Here we give a review on the modular forms of $\Gamma_6$.
The modular forms of level 6 of even weights can be constructed from the products of the Dedekind eta function \cite{Li:2021buv},
\begin{align}
  \eta(\tau) = q^{1/24} \prod_{n=1}^\infty (1-q^n), \quad q = e^{2\pi i\tau}.
\end{align}
Using $\eta$, four linearly independent modular forms of weight 2 can be written down as
\begin{align}
  &Y^{(2)}_{\bm{3^0}}(\tau) = 
  \begin{pmatrix}
    -Y_1^2 \\ \sqrt{2}Y_1Y_2 \\ Y_2^2 \\
  \end{pmatrix}, \quad
  Y^{(2)}_{\bm{1^1_2}}(\tau) = Y_3Y_6-Y_4Y_5, \quad
  Y^{(2)}_{\bm{2_0}}(\tau) = \frac{1}{\sqrt{2}}
  \begin{pmatrix}
    Y_1Y_4-Y_2Y_3 \\ Y_1Y_6-Y_2Y_5 \\
  \end{pmatrix}, \\
  &Y^{(2)}_{\bm{6}}(\tau) = \frac{1}{\sqrt{2}}
  \begin{pmatrix}
    Y_1Y_4+Y_2Y_3 \\ \sqrt{2}Y_2Y_4 \\ -\sqrt{2}Y_1Y_3 \\ Y_1Y_6+Y_2Y_5 \\ \sqrt{2}Y_2Y_6 \\ -\sqrt{2}Y_1Y_5 \\
  \end{pmatrix},
\end{align}
where
\begin{align}
  &Y_1(\tau) = 3\frac{\eta^3(3\tau)}{\eta(\tau)}+\frac{\eta^3(\tau/3)}{\eta(\tau)}, \\
  &Y_2(\tau) = 3\sqrt{2} \frac{\eta^3(3\tau)}{\eta(\tau)}, \\
  &Y_3(\tau) = 3\sqrt{2}\frac{\eta^3(6\tau)}{\eta(2\tau)}, \\
  &Y_4(\tau) = -3\frac{\eta^3(6\tau)}{\eta(2\tau)}-\frac{\eta^3(2\tau/3)}{\eta(2\tau)}, \\
  &Y_5(\tau) = \sqrt{6}\frac{\eta^3(6\tau)}{\eta(2\tau)}-\sqrt{6}\frac{\eta^3(3\tau/2)}{\eta(\tau/2)}, \\
  &Y_6(\tau) = -\sqrt{3}\frac{\eta^3(6\tau)}{\eta(2\tau)}+\frac{1}{\sqrt{3}}\frac{\eta^3(\tau/6)}{\eta(\tau/2)}-\frac{1}{\sqrt{3}}\frac{\eta^3(2\tau/3)}{\eta(2\tau)}+\sqrt{3}\frac{\eta^3(3\tau/2)}{\eta(\tau/2)}.
\end{align}
Then we can construct the modular forms of weight 4 by the CG coefficients shown in appendix \ref{appendix:A} as
\begin{align}
  &Y^{(4)}_{\bm{1^0_0}}(\tau) = \left(Y^{(2)}_{\bm{2_0}} Y^{(2)}_{\bm{2_0}}\right)_{\bm{1^0_0}}, \quad
  Y^{(4)}_{\bm{1^0_1}}(\tau) = \left(Y^{(2)}_{\bm{1^1_2}} Y^{(2)}_{\bm{1^1_2}}\right)_{\bm{1^0_1}}, \quad
  Y^{(4)}_{\bm{2_0}}(\tau) = \left(Y^{(2)}_{\bm{2_0}} Y^{(2)}_{\bm{2_0}}\right)_{\bm{2_0}}, \\
  &Y^{(4)}_{\bm{2_2}}(\tau) = \left(Y^{(2)}_{\bm{1^1_2}} Y^{(2)}_{\bm{2_0}}\right)_{\bm{2_2}}, \quad
  Y^{(4)}_{\bm{3^0}}(\tau) = \left(Y^{(2)}_{\bm{2_0}} Y^{(2)}_{\bm{6}}\right)_{\bm{3^0}}, \quad
  Y^{(4)}_{\bm{3^1}}(\tau) = \left(Y^{(2)}_{\bm{1^1_2}} Y^{(2)}_{\bm{3^0}}\right)_{\bm{3^1}}, \\
  &Y^{(4)}_{\bm{6i}}(\tau) = \left(Y^{(2)}_{\bm{1^1_2}} Y^{(2)}_{\bm{6}}\right)_{\bm{6}}, \quad
  Y^{(4)}_{\bm{6ii}}(\tau) = \left(Y^{(2)}_{\bm{2_0}} Y^{(2)}_{\bm{3^0}}\right)_{\bm{6}}.
\end{align}
Note that $Y^{(4)}_{\bm{6i}}$ and $Y^{(4)}_{\bm{6ii}}$ stand for two linearly independent six-dimensional modular forms of weight 4.
We use the same convention for other modular forms.
Similarly, we construct the modular forms of weight 6 as
\begin{align}
  &Y^{(6)}_{\bm{1^0_0}}(\tau) = \left(Y^{(2)}_{\bm{2_0}} Y^{(4)}_{\bm{2_0}}\right)_{\bm{1^0_0}}, \quad
  Y^{(6)}_{\bm{1^1_0}}(\tau) = \left(Y^{(2)}_{\bm{1^1_2}} Y^{(4)}_{\bm{1^0_1}}\right)_{\bm{1^1_0}}, \quad
  Y^{(6)}_{\bm{1^1_2}}(\tau) = \left(Y^{(2)}_{\bm{1^1_2}} Y^{(4)}_{\bm{1^0_0}}\right)_{\bm{1^1_2}}, \\
  &Y^{(6)}_{\bm{2_0}}(\tau) = \left(Y^{(2)}_{\bm{2_0}} Y^{(4)}_{\bm{1^0_0}}\right)_{\bm{2_0}}, \quad
  Y^{(6)}_{\bm{2_1}}(\tau) = \left(Y^{(2)}_{\bm{2_0}} Y^{(4)}_{\bm{1^0_1}}\right)_{\bm{2_1}}, \quad
  Y^{(6)}_{\bm{2_2}}(\tau) = \left(Y^{(2)}_{\bm{1^1_2}} Y^{(4)}_{\bm{2_0}}\right)_{\bm{2_2}}, \\
  &Y^{(6)}_{\bm{3^0i}}(\tau) = \left(Y^{(2)}_{\bm{3^0}} Y^{(4)}_{\bm{1^0_1}}\right)_{\bm{3^0}}, \quad
  Y^{(6)}_{\bm{3^0ii}}(\tau) = \left(Y^{(2)}_{\bm{3^0}} Y^{(4)}_{\bm{1^0_0}}\right)_{\bm{3^0}}, \quad
  Y^{(6)}_{\bm{3^1}}(\tau) = \left(Y^{(2)}_{\bm{1^1_2}} Y^{(4)}_{\bm{3^0}}\right)_{\bm{3^1}}, \\
  &Y^{(6)}_{\bm{6i}}(\tau) = \left(Y^{(2)}_{\bm{2_0}} Y^{(4)}_{\bm{3^1}}\right)_{\bm{6}}, \quad
  Y^{(6)}_{\bm{6ii}}(\tau) = \left(Y^{(2)}_{\bm{6}} Y^{(4)}_{\bm{1^0_0}}\right)_{\bm{6}}, \quad
  Y^{(6)}_{\bm{6iii}}(\tau) = \left(Y^{(2)}_{\bm{3^0}} Y^{(4)}_{\bm{2_0}}\right)_{\bm{6}}.
\end{align}
In Table \ref{tab:even_weight_modular_forms} we summarize the modular forms of level 6 of even weights up to 6.
\begin{table}[H]
 \centering
 \renewcommand{\arraystretch}{1.4}
  \begin{tabular}{c|c}
   \hline
   & Modular form $Y^{(k_Y)}_{\bm{r}}$ \\ \hline
   $k_Y=2$ & $Y_{\bm{1^1_2}}^{(2)}$, $Y_{\bm{2_0}}^{(2)}$, $Y_{\bm{3^0}}^{(2)}$, $Y_{\bm{6}}^{(2)}$ \\ \hline
   $k_Y=4$ & $Y_{\bm{1^0_0}}^{(4)}$, $Y_{\bm{1^0_1}}^{(4)}$, $Y_{\bm{2_0}}^{(4)}$, $Y_{\bm{2_2}}^{(4)}$, $Y_{\bm{3^0}}^{(4)}$, $Y_{\bm{3^1}}^{(4)}$, $Y_{\bm{6i}}^{(4)}$, $Y_{\bm{6ii}}^{(4)}$ \\ \hline
   $k_Y=6$ & $Y_{\bm{1^0_0}}^{(6)}$, $Y_{\bm{1^1_0}}^{(6)}$, $Y_{\bm{1^1_2}}^{(6)}$, $Y_{\bm{2_0}}^{(6)}$, $Y_{\bm{2_1}}^{(6)}$, $Y_{\bm{2_2}}^{(6)}$, $Y_{\bm{3^0i}}^{(6)}$, $Y_{\bm{3^0ii}}^{(6)}$, $Y_{\bm{3^1}}^{(6)}$, $Y_{\bm{6i}}^{(6)}$, $Y_{\bm{6ii}}^{(6)}$, $Y_{\bm{6iii}}^{(6)}$ \\ \hline
  \end{tabular}
  \caption{The modular forms of level 6 of even weights up to 6.}
  \label{tab:even_weight_modular_forms}
\end{table}
Also we construct the singlet modular forms of weights 8, 10, 12 and 14 which we have used in our analysis.
First, the singlet modular forms of weight 8 are given by
\begin{align}
  Y_{\bm{1^0_0}}^{(8)} = \left(Y_{\bm{1^0_0}}^{(4)} Y_{\bm{1^0_0}}^{(4)}\right)_{\bm{1^0_0}}, \quad
  Y_{\bm{1^0_1}}^{(8)} = \left(Y_{\bm{1^0_0}}^{(4)} Y_{\bm{1^0_1}}^{(4)}\right)_{\bm{1^0_1}}, \quad
  Y_{\bm{1^0_2}}^{(8)} = \left(Y_{\bm{1^0_1}}^{(4)} Y_{\bm{1^0_1}}^{(4)}\right)_{\bm{1^0_2}}, \quad
  Y_{\bm{1^1_2}}^{(8)} = \left(Y_{\bm{2_0}}^{(4)} Y_{\bm{2_2}}^{(4)}\right)_{\bm{1^1_2}}.
\end{align}
The singlet modular forms of weight 10 are given by
\begin{align}
  &Y_{\bm{1^0_0}}^{(10)} = \left(Y_{\bm{1^0_0}}^{(4)} Y_{\bm{1^0_0}}^{(6)}\right)_{\bm{1^0_0}}, \quad
  Y_{\bm{1^0_1}}^{(10)} = \left(Y_{\bm{1^0_1}}^{(4)} Y_{\bm{1^0_0}}^{(6)}\right)_{\bm{1^0_1}}, \quad
  Y_{\bm{1^1_0}}^{(10)} = \left(Y_{\bm{1^0_0}}^{(4)} Y_{\bm{1^1_0}}^{(6)}\right)_{\bm{1^1_0}}, \\
  &Y_{\bm{1^1_1}}^{(10)} = \left(Y_{\bm{1^0_1}}^{(4)} Y_{\bm{1^1_0}}^{(6)}\right)_{\bm{1^1_1}}, \quad
  Y_{\bm{1^1_2}}^{(10)} = \left(Y_{\bm{1^0_0}}^{(4)} Y_{\bm{1^1_2}}^{(6)}\right)_{\bm{1^1_2}}.
\end{align}
The singlet modular forms of weight 12 are given by
\begin{align}
  &Y_{\bm{1^0_0i}}^{(12)} = \left(Y_{\bm{1^0_0}}^{(6)} Y_{\bm{1^0_0}}^{(6)}\right)_{\bm{1^0_0}}, \quad
  Y_{\bm{1^0_0ii}}^{(12)} = \left(Y_{\bm{1^1_0}}^{(6)} Y_{\bm{1^1_0}}^{(6)}\right)_{\bm{1^0_0}}, \quad
  Y_{\bm{1^0_1}}^{(12)} = \left(Y_{\bm{1^1_2}}^{(6)} Y_{\bm{1^1_2}}^{(6)}\right)_{\bm{1^0_1}}, \\
  &Y_{\bm{1^0_2}}^{(12)} = \left(Y_{\bm{1^1_0}}^{(6)} Y_{\bm{1^1_2}}^{(6)}\right)_{\bm{1^0_2}}, \quad
  Y_{\bm{1^1_0}}^{(12)} = \left(Y_{\bm{1^0_0}}^{(6)} Y_{\bm{1^1_0}}^{(6)}\right)_{\bm{1^1_0}}, \quad
  Y_{\bm{1^1_2}}^{(12)} = \left(Y_{\bm{1^0_0}}^{(6)} Y_{\bm{1^1_2}}^{(6)}\right)_{\bm{1^1_2}}.
\end{align}
The singlet modular forms of weight 14 are given by
\begin{align}
  &Y_{\bm{1^0_0}}^{(14)} = \left(Y_{\bm{1^0_0}}^{(6)} Y_{\bm{1^0_0}}^{(8)}\right)_{\bm{1^0_0}}, \quad
  Y_{\bm{1^0_1}}^{(14)} = \left(Y_{\bm{1^0_0}}^{(6)} Y_{\bm{1^0_1}}^{(8)} \right)_{\bm{1^0_1}}, \quad
  Y_{\bm{1^0_2}}^{(14)} = \left(Y_{\bm{1^0_0}}^{(6)} Y_{\bm{1^0_2}}^{(8)} \right)_{\bm{1^0_2}}, \\
  &Y_{\bm{1^1_0}}^{(14)} = \left(Y_{\bm{1^1_0}}^{(6)} Y_{\bm{1^0_0}}^{(8)} \right)_{\bm{1^1_0}}, \quad
  Y_{\bm{1^1_1}}^{(14)} = \left(Y_{\bm{1^1_0}}^{(6)} Y_{\bm{1^0_1}}^{(8)} \right)_{\bm{1^1_1}}, \quad
  Y_{\bm{1^1_2i}}^{(14)} = \left(Y_{\bm{1^1_2}}^{(6)} Y_{\bm{1^0_0}}^{(8)} \right)_{\bm{1^1_2}}, \\
  &Y_{\bm{1^1_2ii}}^{(14)} = \left(Y_{\bm{1^1_0}}^{(6)} Y_{\bm{1^0_2}}^{(8)} \right)_{\bm{1^1_2}}.
\end{align}
In Table \ref{tab:even_weight_singlet_modular_forms} we summarize the singlet modular forms of level 6 of weights 8, 10, 12 and 14.
\begin{table}[H]
 \centering
 \renewcommand{\arraystretch}{1.4}
  \begin{tabular}{c|c}
   \hline
   & Modular form $Y^{(k_Y)}_{\bm{r}}$ \\ \hline
   $k_Y=8$ & $Y_{\bm{1^0_0}}^{(8)}$, $Y_{\bm{1^0_1}}^{(8)}$, $Y_{\bm{1^0_2}}^{(8)}$, $Y_{\bm{1^1_2}}^{(8)}$ \\ \hline
   $k_Y=10$ & $Y_{\bm{1^0_0}}^{(10)}$, $Y_{\bm{1^0_1}}^{(10)}$, $Y_{\bm{1^1_0}}^{(10)}$, $Y_{\bm{1^1_1}}^{(10)}$, $Y_{\bm{1^1_2}}^{(10)}$ \\ \hline
   $k_Y=12$ & $Y_{\bm{1^0_0i}}^{(12)}$, $Y_{\bm{1^0_0ii}}^{(12)}$, $Y_{\bm{1^0_1}}^{(12)}$, $Y_{\bm{1^0_2}}^{(12)}$, $Y_{\bm{1^1_0}}^{(12)}$, $Y_{\bm{1^1_2}}^{(12)}$ \\ \hline
   $k_Y=14$ & $Y^{(14)}_{\bm{1^0_0}}$, $Y^{(14)}_{\bm{1^0_1}}$, $Y^{(14)}_{\bm{1^0_2}}$, $Y^{(14)}_{\bm{1^1_0}}$, $Y^{(14)}_{\bm{1^1_1}}$, $Y^{(14)}_{\bm{1^1_2i}}$, $Y^{(14)}_{\bm{1^1_2ii}}$ \\ \hline
  \end{tabular}
  \caption{The singlet modular forms of level 6 of weights 8, 10, 12 and 14.}
  \label{tab:even_weight_singlet_modular_forms}
\end{table}

\clearpage

\end{document}